\documentclass[sigconf]{acmart} 

\usepackage{algorithmic}
\usepackage[ruled,linesnumbered,vlined]{algorithm2e}
\usepackage{setspace}
\usepackage{listings}
\usepackage{multicol}
\usepackage{multirow}
\usepackage{makecell}
\usepackage{amsthm}
\usepackage{enumitem}
\usepackage{tikz}
\usepackage{xspace}
\usepackage{wrapfig}
\usepackage{graphicx,calc, scalerel}
\usepackage{mathtools}
\usepackage{caption}
\usepackage{subcaption}
\usepackage{xcolor, colortbl}
\usepackage{bbm}
\usepackage{pifont}

\copyrightyear{2024}
\acmYear{2024}
\setcopyright{acmlicensed}
\acmConference[ICSE '24]{2024 IEEE/ACM 46th International Conference on Software Engineering}{April 14--20, 2024}{Lisbon, Portugal}
\acmBooktitle{2024 IEEE/ACM 46th International Conference on Software Engineering (ICSE '24), April 14--20, 2024, Lisbon, Portugal}
\acmDOI{10.1145/3597503.3639121}
\acmISBN{979-8-4007-0217-4/24/04}

\newcommand{\Comment}[1]{}

\newcommand{\CodeIn}[1]{{\small \texttt{#1}}}
\newcommand{\parabf}[1]{\noindent\textbf{#1}}

\newcommand{\tech}{\textsc{Fuzz4All}\xspace} %
\newcommand{\distillation}{distillation\xspace}
\newcommand{\generation}{generation\xspace}
\newcommand{\mutate}{\CodeIn{mutate-existing}\xspace}
\newcommand{\generatenew}{\CodeIn{generate-new}\xspace}
\newcommand{\semanticequiv}{\CodeIn{semantic-equiv}\xspace}
\newcommand{\all}{\CodeIn{all}\xspace}

\newcommand{\noinput}{\CodeIn{no input}\xspace}
\newcommand{\documentation}{\CodeIn{raw prompt}\xspace}
\newcommand{\autoprompt}{\CodeIn{autoprompt}\xspace}
\newcommand{\smallautoprompt}{\CodeIn{\footnotesize autoprompt}\xspace}
\newcommand{\noloop}{\CodeIn{w/o example}\xspace}
\newcommand{\addexample}{\CodeIn{w/ example}\xspace}
\newcommand{\smalladdexample}{\CodeIn{\footnotesize w/ example}\xspace}

\newcommand{\totalbugs}{98\xspace}
\newcommand{\totalconfirmedbugs}{64\xspace}
\newcommand{\coverageimprove}{36.8}

\newcommand{\llm}{LLM\xspace}
\newcommand{\llmfull}{large language model\xspace}
\newcommand{\sut}{SUT\xspace}
\newcommand{\nlp}{NLP\xspace}
\newcommand{\nlpfull}{natural language processing\xspace}

\newcommand{\gpt}{GPT\xspace}
\newcommand{\starcoder}{StarCoder\xspace}
\newcommand{\bert}{BERT\xspace}
\newcommand{\codebert}{CodeBERT\xspace}
\newcommand{\bart}{BART\xspace}
\newcommand{\codetf}{CodeT5\xspace}
\newcommand{\chatgpt}{ChatGPT\xspace}
\newcommand{\codex}{Codex\xspace}
\newcommand{\incoder}{InCoder\xspace}

\newcommand{\clang}{Clang\xspace}
\newcommand{\gcc}{GCC\xspace}
\newcommand{\zee}{Z3\xspace}
\newcommand{\cvcf}{CVC5\xspace}
\newcommand{\qiskit}{Qiskit\xspace}

\newcommand{\afl}{\textsc{AFL}\xspace}
\newcommand{\libfuzzer}{\textsc{libFuzzer}\xspace}
\newcommand{\polyglot}{\textsc{PolyGlot}\xspace}
\newcommand{\jsfunfuzz}{\textsc{jsfunfuzz}\xspace}

\newcommand{\syzkaller}{\textsc{Syz\-kaller}\xspace}
\newcommand{\csmith}{\textsc{Csmith}\xspace}
\newcommand{\yarpgen}{\textsc{YARPGen}\xspace}
\newcommand{\grayc}{\textsc{GrayC}\xspace}
\newcommand{\typefuzz}{\textsc{TypeFuzz}\xspace}
\newcommand{\gofuzz}{\textsc{go-fuzz}\xspace}
\newcommand{\morphq}{\textsc{MorphQ}\xspace}
\newcommand{\hephaestus}{\textsc{Hephaestus}\xspace}
\newcommand{\treefuzz}{\textsc{TreeFuzz}\xspace}
\newcommand{\titanfuzz}{\textsc{TitanFuzz}\xspace}

\newcommand{\codemosa}{\textsc{CodaMosa}\xspace}
\newcommand{\testpilot}{\textsc{TestPilot}\xspace}

\newcommand{\eg}{e.g.,\xspace}
\newcommand{\ie}{i.e.,\xspace}

\DeclareMathOperator*{\argmax}{arg\,max}
\newcommand*{\argmaxl}{\argmax\limits}

\newcommand*\circled[1]{\scalebox{0.8}{\tikz[baseline=(char.base)]{
\node[anchor=text, shape=circle,fill, inner sep=0pt, minimum size=1.2em] (char) {\footnotesize \textcolor{white}{#1}};}}}

\definecolor{codegreen}{rgb}{0,0.6,0}
\definecolor{codegray}{rgb}{0.5,0.5,0.5}
\definecolor{codepurple}{rgb}{0.58,0,0.82}
\definecolor{backcolour}{rgb}{0.95,0.95,0.92}
\definecolor{mygreen}{rgb}{1, 0, 0.6}

\lstdefinestyle{mystyle}{
    backgroundcolor=\color{backcolour},
    commentstyle=\color{codegreen},
    keywordstyle=\color{magenta},
    numberstyle=\tiny\color{codegray},
    stringstyle=\color{codepurple},
    basicstyle=\ttfamily\footnotesize,
    breakatwhitespace=false,
    breaklines=true,
    captionpos=b,
    keepspaces=true,
    numbers=left,
    numbersep=5pt,
    showspaces=false,
    showstringspaces=false,
    showtabs=false,
    tabsize=2
}
\makeatletter
\lst@AddToHook{PreSet}{\normallineskiplimit=0pt}
\makeatother
\begin{document}

\title{\tech: Universal Fuzzing with Large Language Models} %

\author{Chunqiu Steven Xia}
    \affiliation{\institution{University of Illinois\\ Urbana-Champaign, USA}\country{}}
    \email{chunqiu2@illinois.edu}
\author{Matteo Paltenghi}
    \affiliation{\institution{University of\\ Stuttgart, Germany}\country{}}
    \email{mattepalte@live.it}
\author{Jia Le Tian}
    \affiliation{\institution{University of Illinois\\ Urbana-Champaign, USA}\country{}}
    \email{jialelt2@illinois.edu}
\author{Michael Pradel}
    \affiliation{\institution{University of\\ Stuttgart, Germany}\country{}}
    \email{michael@binaervarianz.de}
\author{Lingming Zhang}
    \affiliation{\institution{University of Illinois\\ Urbana-Champaign, USA}\country{}}
    \email{lingming@illinois.edu}

\begin{abstract}

Fuzzing has achieved tremendous success in discovering bugs and vulnerabilities in various software systems.
Systems under test (\sut{s}) that take in programming or formal language as inputs, \eg compilers, runtime engines, constraint solvers, and software libraries with accessible APIs, are especially important as they are fundamental  building blocks of software development.
However, existing fuzzers for such systems often target a specific language, and thus cannot be easily applied to other languages or even other versions of the same language.
Moreover, the inputs generated by existing fuzzers are often limited to specific features of the input language, and thus can hardly reveal bugs related to other or new features.
This paper presents \tech, the first fuzzer that is \emph{universal} in the sense that it can target many different input languages and many different features of these languages.
The key idea behind \tech is to leverage \llmfull{s} (\llm{s}) as an input generation and mutation engine, which enables the approach to produce diverse and realistic inputs for any practically relevant language.
To realize this potential, we present a novel autoprompting technique, which creates \llm prompts that are well-suited for fuzzing, and a novel \llm-powered fuzzing loop, which iteratively updates the prompt to create new fuzzing inputs.
We evaluate \tech on nine systems under test that take in six different languages (C, C++, Go, SMT2, Java, and Python) as inputs.
The evaluation shows, across all six languages, that universal fuzzing achieves higher coverage than existing, language-specific fuzzers.
Furthermore, \tech has identified \totalbugs bugs in widely used systems, such as GCC, Clang, Z3, CVC5, OpenJDK, and the Qiskit quantum computing platform, with \totalconfirmedbugs bugs already confirmed by developers as previously unknown.

\end{abstract}
\maketitle

\section{Introduction}

Fuzz testing~\cite{SuttonFuzzingBook, zeller2019fuzzing}, also known as fuzzing, is an automated testing approach for generating inputs designed to expose unexpected behaviors, \eg crashes, of a system under test (\sut).
Researchers and practitioners have successfully built practical fuzzing tools, which have shown great success in finding numerous bugs and vulnerabilities in real-world systems~\cite{bohme2020fuzzing}.
A particularly important family of \sut{s} are systems that take in programming or formal language inputs, \eg compilers, runtime engines, and constraint solvers.
Numerous fuzzers have been proposed for such systems since they are the fundamental building blocks for software development~\cite{chen2020survey}.
For example, finding bugs in compilers and runtime engines is crucial because they can affect all corresponding downstream applications.%

Traditional fuzzers can be categorized into generation-based~\cite{yang2011csmith, jsfunfuzz, livinskii2020yarpgen} and mutation-based~\cite{SuttonFuzzingBook, holler2012fuzzing, evenmendoza2023grayc}.
Generation-based fuzzers aim to directly synthesize complete code snippets, \eg using a pre-defined grammar for the target language.
Instead of synthesizing from scratch, mutation-based fuzzers apply mutation operators or transformation rules to a set of high quality fuzzing seeds.
Unfortunately, both traditional fuzzing approaches face the following limitations and challenges:

\textit{C1: Tight coupling with target system and language.}
Traditional fuzzers are often designed to target a specific language or a particular \sut.
However, designing and implementing a fuzzer is extremely time-consuming.
For example, \csmith~\cite{yang2011csmith}, a fuzzer for C/C++ compilers, has more than 80k lines of code, while \syzkaller~\cite{syzkaller}, a fuzzer for Linux system calls, contains tens of thousands of handcrafted rules~\cite{bulekovno2023grammar} to generate and modify system calls.
Because each target language is different, it is often non-trivial to reuse the effort of implementing a fuzzer from one input language for another.
Furthermore, fuzzing strategies that work well for one \sut may not work at all for another one.

\textit{C2: Lack of support for evolution.}
Real-world systems are constantly evolving, \eg by adding new features to the input language.
Traditional fuzzers designed for a specific version of a language or \sut may lose their effectiveness on a new version and cannot be easily used to test newly implemented features.
For example, \csmith supports only a limited set of features up to C++11, while the C++ language has evolved significantly since then.
In fact, recent work~\cite{even2022csmithedge} shows that over a six-month fuzzing period, \csmith was not able to uncover any new bugs in the latest releases of the \gcc and \clang compilers, showing that new versions of compilers are becoming immune to existing fuzzers.

\textit{C3: Restricted generation ability.}
Even within the scope of a specific target language, both generation-based and mutation-based fuzzing often are unable to cover a large part the input space. 
Generation-based fuzzers heavily rely on an input grammar to synthesize valid code, and additionally are equipped with semantic rules that ensure the validity of the synthesized code.
To generate a high amount of valid fuzzing inputs or to side-step difficult-to-model language features, generation-based fuzzers often use a subset of the full language grammar, which limits them to test only a subset of all language features.
Similarly, mutation-based fuzzers are limited by their mutation operators and require high quality seeds that can be difficult to obtain.

\parabf{Our work.}
We present \tech, the first fuzzer that is \emph{universal} in the sense that it can target many different input languages and many different features of theses languages.
Our approach fundamentally differs from existing general-purpose fuzzers, \eg \afl~\cite{afl} and \libfuzzer~\cite{libfuzzer}, which use extremely simple mutations, are unaware of the target language, and therefore struggle to produce meaningful programming language fuzzing inputs.
Instead, our key idea is to leverage a \llmfull (\llm) as an input generation and mutation engine.
Because \llm{s} are pre-trained on large amounts of examples in various programming languages and other formal languages, they come with an implicit understanding of the syntax and semantics of these languages.
\tech leverages this ability by using an \llm as a universal input generation and mutation engine.

The input to \tech are user-provided documents describing the SUT, and optionally, specific features of the SUT to focus on, \eg in the form of documentation, example code, or formal specifications.
However, these user inputs may be too verbose to directly use as a prompt for the \llm.
Instead of requiring the user to manually engineer a prompt~\cite{Liu2021a}, which is time-consuming, we present an \emph{autoprompting} step that automatically distills all user-provided inputs into a concise and effective prompt for fuzzing.
This prompt is the initial input to an \llm that generates fuzzing inputs.
Since continuously sampling with the same prompt would lead to many similar fuzzing inputs, we present an \emph{\llm-powered fuzzing loop}, which iteratively updates the prompt to generate a diverse set of fuzzing inputs.
To this end, \tech combines fuzzing inputs generated in previous iterations with natural language instructions, \eg asking to mutate these inputs.
The \llm-generated fuzzing inputs are then passed to the SUT, which we validate against a user-provided test oracle, such as checking for system crashes.

\tech addresses the previously discussed limitations and challenges of traditional fuzzers.
Instead of meticulously designing a single-purpose fuzzer for a specific \sut (C1), \tech, by using an \llm as the generation engine, can be applied to a wide range of \sut{s} and input languages.
Compared to existing fuzzers that target a specific version of the \sut or input language (C2), \tech can easily evolve with the target.
For example, to fuzz-test a newly implemented feature, a user can simply provide documentation or example code related to that feature.
To address the restricted generation ability of traditional fuzzers (C3), \tech exploits the fact that \llm{s} are pre-trained on billions of code snippets, enabling them to create a wide range of examples that likely obey the syntactic and semantic constraints of the input language.
Finally, \tech does not require any instrumentation of the SUT, making the approach easily applicable in practice.

We perform an extensive evaluation on six input languages (C, C++, SMT, Go, Java, and Python) and nine \sut{s}.
For each of them, we compare our approach against state-of-the-art generation-based and mutation-based fuzzers.
The results show that \tech achieves the highest code coverage across all languages, improving the previous state-of-the-art coverage by \coverageimprove\%, on average.
Additionally, we demonstrate that \tech supports both general fuzzing and fuzzing targeted at specific features of the \sut, which a user decides upon by providing adequate input documents.
Finally, \tech detects \totalbugs bugs across our studied \sut{s}, with \totalconfirmedbugs already confirmed by developers as previously unknown.

\parabf{Contributions:} This paper makes the following contributions:

\begin{itemize}[noitemsep, leftmargin=*, topsep=0pt]
\item[$\star$] \textbf{Universal fuzzing}.
We introduce a new dimension for fuzzing that directly leverages the multi-lingual capabilities of \llm{s} to fuzz-test many \sut{s} with a wide range of meaningful inputs.

\item[$\star$] \textbf{Autoprompting for fuzzing}.
We present a novel autoprompting stage to support both general and targeted fuzzing by automatically distilling user inputs into a prompt that is effective at generating inputs to the \sut.

\item[$\star$] \textbf{\llm-powered fuzzing loop}.
We present an algorithm that continuously generates new fuzzing inputs by iteratively modifying the prompt with selected examples and generation strategies.

\item[$\star$] \textbf{Evidence of real-world effectiveness}.
We show across six popular languages and nine real-world \sut{s} (e.g., GCC, CVC5, Go, javac, and Qiskit) that our approach significantly improves coverage compared to state-of-the-art fuzzers (avg. \coverageimprove{}\%) and detects \totalbugs bugs, with \totalconfirmedbugs already confirmed as previously unknown.

\end{itemize}
\section{Background and Related Work}

\subsection{Large Language Models}

Recent developments in \nlpfull (\nlp)
has lead to the wide-spread adoption of \llmfull{s} (\llm{s}) for both natural language~\cite{brown2020gpt3} and code tasks~\cite{xu2022systematic}.
State-of-the-art \llm{s} are based on transformers~\cite{vaswani2017attention} and can be classified into decoder-only (\eg \gpt{3}~\cite{brown2020gpt3} and \starcoder~\cite{li2023starcoder}), encoder-only (\eg \bert~\cite{devlin2018bert} and \codebert~\cite{feng2020codebert}) and encoder-decoder (\bart~\cite{lewis2019bart} and \codetf~\cite{wang2021codet5}) models.
More recently, instruction-based \llm{s} (\eg \chatgpt~\cite{chatgpt} and \gpt{4}~\cite{openai2023gpt4}) and \llm{s} fine-tuned using reinforcement learning from human feedback (RLHF)~\cite{ziegler2019rlhf} are shown to understand and follow complex instructions~\cite{ouyang2022instructgpt, chatgpt, bang2023multitask}.

\llm{s} are typically either fine-tuned~\cite{radford2018improving} or prompted~\cite{Liu2021a} to perform specific tasks.
Fine-tuning updates the model weights through further training on a task-specific dataset.
However, suitable datasets may be unavailable, and as \llm sizes continue to grow~\cite{kaplan2020scaling}, fine-tuning an \llm is also increasingly expensive.
Prompting, on the other hand, does not require explicitly updating the model weights, but provides the \llm with a description of the task, and optionally, a few examples of solving the task.
The process of picking the input (\ie prompt) is known as prompt engineering~\cite{Liu2021a}, where a user tries different input instructions until finding one that works well.
Recently, researchers have proposed \textit{autoprompting}~\cite{shin2020autoprompt}, an automatic process that uses \llm gradients to select either soft prompts~\cite{qin2021learning, li2021prefix}, i.e., continuous vector embeddings, or hard prompts~\cite{tam2021improving, schick2020exploiting}, i.e., natural language text.
Even more recently, researchers have substituted gradient-based methods by computing a proxy score of effectiveness~\cite{zhou2022humanprompt}.

This work leverages \llm{s} for the important problem of fuzzing. Unlike traditional autoprompting and proxy-based approaches, our autoprompting strategy directly synthesizes prompts using \gpt{4} and scores them according to a fuzzing-specific goal.%

\subsection{Fuzzing and Testing}

Fuzz testing aims to generate inputs that cause unexpected behaviors of the \sut.
Traditional fuzzers can be classified into generation-based~\cite{yang2011csmith, jsfunfuzz, livinskii2020yarpgen} and mutation-based~\cite{SuttonFuzzingBook, holler2012fuzzing, evenmendoza2023grayc}.
Generation-based fuzzers create complete code snippets using pre-defined grammars and built-in knowledge of the semantics of the target language.
\csmith~\cite{yang2011csmith} and \yarpgen~\cite{livinskii2020yarpgen} hard-code language specifications to ensure the validity of generated code snippets to test C and C++ compilers, respectively.
\jsfunfuzz~\cite{jsfunfuzz} combines a language grammar with historical bug-triggering code snippets to generate new inputs to test JavaScript engines.
Generation-based fuzzers have also been used to test OpenCL~\cite{lidbury2015clsmith}, the JVM~\cite{chaliasos2022hephaestus}, CUDA~\cite{jiang2020cudasmith}, deep learning compilers~\cite{liu2023nnsmith}, Datalog engines~\cite{DBLP:conf/sigsoft/MansurCW21}, and interactive debuggers~\cite{fse2018}.
Mutation-based fuzzers~\cite{SuttonFuzzingBook} iteratively perform transformations on seeds to generate new fuzzing inputs.
In addition to basic mutations, researchers have developed complex transformations targeted at ensuring type consistency~\cite{chaliasos2022hephaestus, park2021generative}, adding historical bug-triggering code snippets~\cite{holler2012fuzzing, zhao2022javatailor}, and coverage feedback~\cite{evenmendoza2023grayc, aschermann2019nautilus, liu2022coverage}.
To benefit from both generation and mutation, many fuzzers use a combination of both approaches~\cite{chen2020survey, ma2023survey}.

Different from the above fuzzers, which target specific \sut{s} or languages, another line of research is on general-purpose fuzzing.
\afl~\cite{afl} and \libfuzzer~\cite{libfuzzer} are general-purpose fuzzers that use genetic algorithms with a fitness function to prioritize fuzzing inputs for further mutations that achieve new coverage.
These mutations are unaware of the \sut{} and focus on byte-level transformations.
That is, when applied on \sut{s} that receive programming languages as input, general-purpose fuzzers are extremely unlikely to produce valid inputs.
Recent work~\cite{groce2022making} has instead added regular expression-based mutation operators to match common programming statements (\eg change \CodeIn{+} to \CodeIn{-}).
The simplicity of these mutation operators limits the ability of such fuzzers at covering new code, especially in more complex languages, such as C~\cite{groce2022making, evenmendoza2023grayc}.
\polyglot~\cite{chen2021polyglot} is another language-agnostic fuzzer, which first parses the seed programs into a uniform intermediate representation using a language-specific grammar and then uses a set of mutation operators to generate new programs.
While promising, \polyglot{} still uses a limited set of mutations and cannot achieve the same level of coverage as fuzzers that are designed for a particular language~\cite{evenmendoza2023grayc}.

To complement traditional fuzzing techniques and apply fuzzing to emerging domains, learning-based fuzzers have been proposed.
Prior learning-based techniques mainly focus on training a neural network to generate fuzzing inputs.
\treefuzz~\cite{patra2016learning} parses the training corpus into a tree structure and through tree traversal, learns a probabilistic, generative model that synthesizes new fuzzing inputs.
Deep learning models have been used to fuzz PDF parsers~\cite{godefroid2017learn}, OpenCL~\cite{cummins2018compiler}, C~\cite{liu2019deepfuzz}, network protocols~\cite{seqfuzzer}, and JavaScript~\cite{lee2020montage}.
Very recently, researchers have also directly leveraged \llm{s} for fuzzing specific libraries, \eg
\titanfuzz~\cite{deng2023titanfuzz} uses \codex~\cite{codex} to generate seed programs and \incoder~\cite{incoder} to perform template-based mutation for fuzzing deep learning libraries~\cite{PyTorch, Tensorflow}.

Unlike prior learning- and \llm-based fuzzers, \tech is easily applicable across many programming languages. Prior work trains language-specific models or requires language-specific parsing. Even \titanfuzz, a recent \llm-based approach, is designed specifically for deep learning libraries with hand-crafted prompts and mutation patterns, and therefore cannot be easily extended to other \sut{s}.
Furthermore, unlike existing techniques, which produce general fuzzing inputs in a particular language, \tech additionally supports targeted fuzzing, which can generate code snippets that focus on selected features.

In addition to fuzzing, \llm{s} have also been applied to the related problem of unit test generation~\cite{lemieux2023codamosa, schafer2023testpilot, nie2023teco, vikram2023propertytest, codexStudy2022, yuan2023manual}. \codemosa~\cite{lemieux2023codamosa} interleaves traditional search-based software testing with querying \codex to generate new unit tests whenever a coverage plateau is reached. \testpilot~\cite{schafer2023testpilot} prompts \codex with method source code and example usages to generate unit tests and to fix incorrectly generated tests.
In contrast to these \llm-based test generators, which require a specific type of input (\eg function source code) and only work for unit testing~\cite{nie2023teco, schafer2023testpilot}, by using our novel autoprompting stage, \tech can take inputs in arbitrary formats for both general and targeted fuzzing.
Furthermore, such unit test generators often require manual work to check or complete the tests as they are limited by automatically generated test-oracles, which even state-of-the-art \llm{s}~\cite{chowdhery2022palm, chatgpt} cannot always produce reliably. %
Instead, \tech leverages widely-used fuzzing oracles, such as crashes, and is fully automated.

\section{\tech Approach}

\begin{figure*}[t]
    \captionsetup{justification=centering}
    \centering
    \includegraphics[keepaspectratio=true,width=0.95\textwidth]{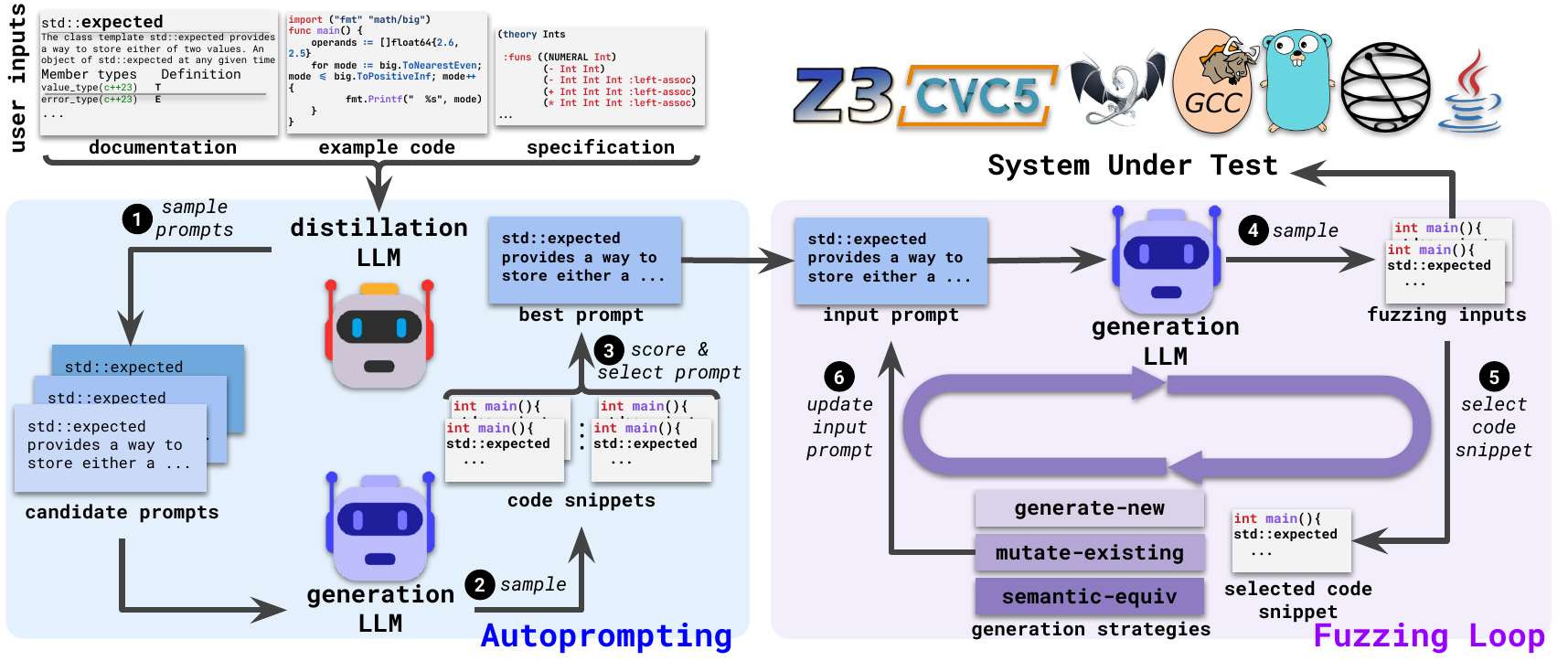}
    \caption{Overview of \tech.
    }
    \label{fig:overview}
\end{figure*}

We present \tech, a universal fuzzer that leverages \llm{s} to support both general and targeted fuzzing of any \sut{s} that take in programming language input.
Figure~\ref{fig:overview} provides an overview of our approach.
\tech first takes in arbitrary \emph{user input} that describes the fuzzing inputs to be generated, \eg documentation of the \sut{}, example code snippets, or specifications.
As the user input may be long, redundant, and partially irrelevant, the approach distills it into a concise but informative prompt for fuzzing.
To this end, \tech performs an \emph{autoprompting} step (Section~\ref{sec:autoprompting}) by using a large, state-of-the-art \emph{\distillation \llm} to sample multiple different candidate prompts \circled{1}.
Each candidate prompt is passed on to the \emph{\generation \llm} to generate code snippets (\ie fuzzing inputs) \circled{2}.
\tech then selects the prompt that produces the highest quality fuzzing inputs \circled{3}.

\tech builds on two models, a \distillation \llm that reduces the given user input and a \generation \llm that creates the fuzzing inputs, to balance the trade-off between the costs and benefits different \llm{s} provide.
Because the \distillation \llm needs to understand and distill arbitrary user input, we use a high-end, large foundational model with strong natural language understanding abilities.
However, directly using such a large model for input generation would be inefficient due to the high inference cost of autoregressive generation.
Instead,  to perform efficient fuzzing, \tech uses a smaller model as the \generation \llm.
While our approach is general across any pairs of \distillation and \generation \llm{s}, we implement \tech with the state-of-the-art \gpt{4}~\cite{openai2023gpt4} and \starcoder~\cite{li2023starcoder}.

Using the best prompt selected via autoprompting as the initial input prompt for the \generation \llm, we then move on to the \emph{fuzzing loop} (Section~\ref{sec:fuzzing_loop}), where \tech continuously samples the \generation \llm to generate fuzzing inputs \circled{4}.
To avoid generating many similar fuzzing inputs, \tech continuously updates the input prompt in each iteration.
Specifically, the approach selects a previously generated input as an \emph{example}~\circled{5}, which demonstrates the kind of future inputs we want the model to generate.
In addition to the example, \tech also appends a \emph{generation instruction} to the initial prompt, which guides the model toward generating new fuzzing inputs \circled{6}.
This process is repeated while continuously passing the generated fuzzing inputs into the \sut and checking its behavior against a user-defined oracle, such as crashes.

\subsection{Autoprompting}
\label{sec:autoprompting}

\begin{algorithm}[t!]
\setstretch{0.6}
\small
\caption{Autoprompting for fuzzing}
\label{alg:autoprompt}
\SetKwData{model}{$\mathcal{M_{D}}$}
\SetKwData{lmodel}{$\mathcal{M_{G}}$}
\SetKwFunction{system}{\sut}
\SetKwData{nsamples}{numSamples}
\SetKwData{inputprompt}{inputPrompt}
\SetKwData{autoinstruction}{APInstruction}
\SetKwData{userinput}{userInput}

\SetKwData{greedyprompt}{greedyPrompt}
\SetKwData{prompts}{candidatePrompts}
\SetKwData{prompt}{prompt}
\SetKwData{p}{p}

\SetKwProg{Fn}{Function}{:}{}
\SetKwFunction{score}{Scoring}
\SetKwFunction{Autoprompting}{Autoprompting}
\SetKwFunction{ConstructPrompt}{ConstructPrompt}
\SetKwInOut{Input}{Input}
\SetKwInOut{Output}{Output}
\SetKw{Break}{break}

\Fn{\Autoprompting}{
    \Input{\userinput, \nsamples }
    \Output{\inputprompt}
    \BlankLine

    \greedyprompt $\leftarrow$ \model(\userinput, \autoinstruction, temp=0) \label{ts:greedy}\\
    \prompts $\leftarrow$ [ \greedyprompt] \\
    \While{$|$\prompts$|$ < \nsamples}{
        \prompt $\leftarrow$ \model(\userinput, \autoinstruction, temp=1) \label{ts:sample}\\
        \prompts $\leftarrow$  \prompts + [ \prompt]  \label{ts:append}\\
    }
    \inputprompt $\leftarrow$ $\argmaxl_{\p\in\prompts}$\score(\lmodel(\p), \system) \label{ts:score}\\
    \algorithmicreturn{ \inputprompt }
}
\end{algorithm}

The following presents the details of the first of two main steps of \tech, which distills the given user input via autoprompting into a prompt suitable for fuzzing.
The user input may describe the \sut in general, or particular feature of the \sut to be tested.
As shown in Figure~\ref{fig:overview}, user inputs may include technical documentation, example code, specifications, or even combinations of different modalities.
Unlike traditional fuzzers that require inputs to follow a specific format, \eg code snippets to use as seeds or well-formed specifications, \tech can directly understand the natural language descriptions or code examples in the user input.
However, some information in the user input may be redundant or irrelevant, and hence, directly using the user inputs as a prompt for the \generation \llm may be ineffective, as confirmed by our ablation study in Section~\ref{sec:ablation_study}.
Therefore, the goal of autoprompting is to generate a distilled input prompt that enables effective \llm-based fuzzing.

\newcommand{\modell}{\mathcal{M_{D}}}
\newcommand{\userinput}{\CodeIn{userInput}}
\newcommand{\instruction}{\CodeIn{APInstruction}}
\newcommand{\prompt}{\CodeIn{prompt}}

\subsubsection{Autoprompting Algorithm}
\label{sec:autoprompt_algorithm}

Algorithm~\ref{alg:autoprompt} details \tech's autoprompting step.
The inputs are the user input and the number of candidate prompts to generate.
The final output is the input prompt selected to be used for the fuzzing campaign.
As our goal is to use a \distillation \llm to generate prompts that distill the information provided by the user, we give the following autoprompting instruction to the \distillation \llm: ``Please summarize the above information in a concise manner to describe the usage and functionality of the target''. Let $\modell$ be the \distillation \llm, \userinput{} be the user input and \instruction{} be the autoprompting instruction. The prompt \prompt{} generated can be formalized as the conditional probability: $\modell (\prompt\; |\; \userinput,\; \instruction)$

\tech first generates a candidate prompt using greedy sampling with temperature $0$ (line~\ref{ts:greedy}).
By first sampling with low temperature, the algorithm obtains a plausible solution with a high degree of confidence.
This approach is commonly used in other domains, \eg program synthesis~\cite{codex}, where the greedy output is evaluated first to check if it can solve the problem.
The algorithm then moves on to sampling with higher temperature to obtain more diverse prompts (line~\ref{ts:sample}), as done in prior work~\cite{codex, xia2023chatrepair}.
Compared to a greedy approach, sampling with high temperature yields different prompts that can each provide a unique distilled summary of the user input.
Each generated prompt is added to a list of candidate prompts (line~\ref{ts:append}), until the algorithm reaches the desired number of candidates.

\newcommand{\modelg}{\mathcal{M_{G}}}
\newcommand{\isvalid}{\CodeIn{isValid}}
\newcommand{\candprompt}{\CodeIn{p}}
\newcommand{\SUT}{\CodeIn{SUT}}
\newcommand{\code}{\CodeIn{c}}
\newcommand{\indf}[1]{\mathbbm{1}_{#1}}
\newcommand{\expf}[2]{\sum_{#1}[#2]}

To pick the best input prompt to be used in the fuzzing step, the algorithm evaluates each candidate prompt by performing a small-scale fuzzing experiment.
Specifically, the approach uses each prompt as an input to the \generation \llm to produce multiple code snippets per prompt.
\tech then scores the generated code snippets for each prompt based on a scoring function.
While the scoring function can be based on a variety of different metrics, \eg coverage, bug finding, or the complexity of generated fuzzing inputs, to make the approach lightweight and general, our scoring function is the number of unique generated code snippets that are valid, i.e., accepted by the target \sut.
This metric is chosen since for fuzzing, we want fuzzing inputs to be valid or close to valid to the logic deep inside the \sut. 
Let $\modelg$ be the generation \llm, \candprompt{} be a candidate prompt, \isvalid{} be a function that returns 1 if a generated code \code{} is valid and 0 otherwise. Our default scoring function is defined as: $\expf{\code \in \modelg(\candprompt)}{\isvalid(\code,~\SUT)}$.
Finally, \tech selects the input prompt with the highest score (line~\ref{ts:score}) as the initial input prompt to be used for fuzzing.
In summary, our autoprompting step combines both prompt generation and scoring, which allows \tech to automatically generate and select a prompt suitable for the fuzzing target.

\subsubsection{Example: Autoprompting}

\begin{figure}[t]
    \captionsetup{justification=centering}
    \centering
    \includegraphics[keepaspectratio=true,width=\linewidth]{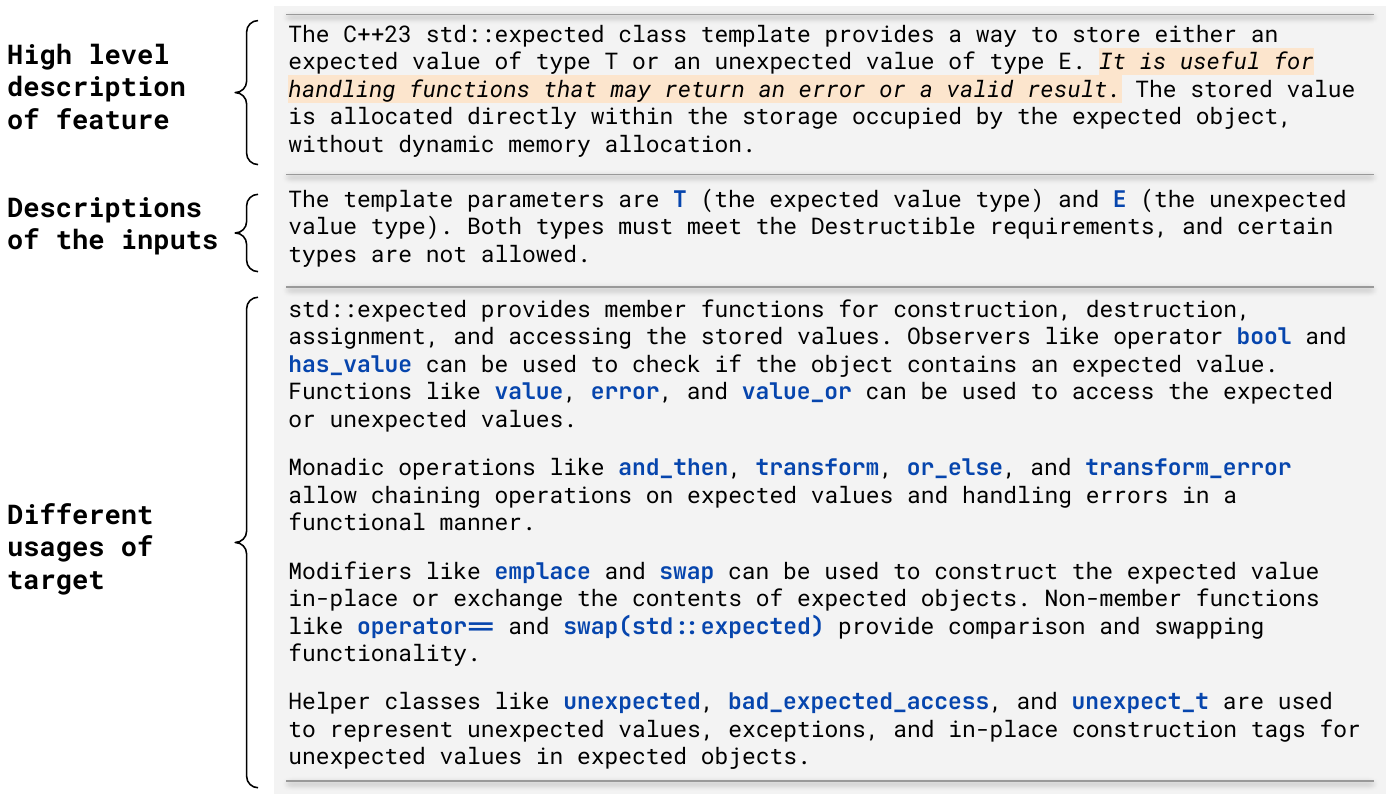}
    \caption{Autoprompting result for \CodeIn{std::expected}.}
    \label{fig:autoprompt_example}
\end{figure}

Figure~\ref{fig:autoprompt_example} shows an example of an input prompt generated by our autoprompting algorithm.
The example is for fuzzing C++ compilers while focusing specifically on \CodeIn{std::expected}, a new feature introduced in C++23.
As the user input, we pass the original \CodeIn{cppreference} documentation~\cite{cppexpecteddocumentation} to \tech, which spans multiple screen lengths with small tables and verbose descriptions (498 words, 3,262 characters).
In contrast, the distilled input prompt created by the autoprompting algorithm provides a more concise natural language description of the targeted feature (214 words, 1,410 characters).
The input prompt contains a high-level description of how \CodeIn{std::expected} is to be used.
For example, the input prompt contains a concise sentence (highlighted in orange) that summarizes the situations the feature is useful in.
Additionally, the input prompt contains descriptions of the inputs, as well as the different usages (\ie member functions) of the feature.
For example, functions \CodeIn{and\_then}, \CodeIn{transform}, \CodeIn{or\_else}, and \CodeIn{transform\_error} have very similar descriptions in the original documentation, which is repeated for each function.
Instead, in the distilled input prompt, these functions are grouped together in a concise manner that still illustrates how they can be used.
Using the distilled input prompt, \tech can generate fuzzing inputs that effectively target the \CodeIn{std::expected} feature of C++ compilers.

\subsubsection{Comparison with Existing Autoprompting Techniques}
To the best of our knowledge, we are the first to automatically distill knowledge from arbitrary user inputs for a software engineering task using black-box autoprompting.
Compared to prior work on autoprompting in NLP~\cite{shin2020autoprompt} and software engineering~\cite{wang2022noprompt}, which optimize the prompt by accessing model gradients, our autoprompting needs only black-box, sampling access to the \distillation \llm.
While the use of a scoring function to evaluate each prompt is similar to recent work in \nlp~\cite{zhou2022humanprompt}, our scoring function directly evaluates the prompt on the exact downstream task of generating valid code snippets, instead of using an approximate proxy scoring function.

\subsection{Fuzzing Loop}
\label{sec:fuzzing_loop}

\newcommand{\modelG}{\mathcal{M_{G}}}
\newcommand{\initialprompt}{P_{i}}
\newcommand{\instructionnew}{I_{n}}
\newcommand{\codesnippet}{c}

Given the input prompt created in the first step of \tech, the goal of the fuzzing loop is to generate diverse fuzzing inputs using a \generation \llm.
However, due to the probabilistic nature of \llm{s}, sampling multiple times using the same input would produce the same or similar code snippets.
For fuzzing, we aim to avoid such repeated inputs and instead want to generate a diverse set of fuzzing inputs that cover new code and discover new bugs.
To accomplish this goal, we exploit the ability of \llm{s} to utilize both examples and natural language instructions to guide the generation.

The high-level idea of the fuzzing loop is to continuously augment the original input prompt by selecting an example fuzzing input from previous iterations and by specifying a generation strategy.
The goal of using an example is to demonstrate the kind of code snippet we want the \generation \llm to produce.
The generation strategies are designed as instructions on what to do with the provided code example.
These strategies are inspired by traditional fuzzers, mimicking their ability to synthesize new fuzzing inputs (as in generation-based fuzzers) and to produce variants of previously generated inputs (as in mutation-based fuzzers).
Before each new iteration of the fuzzing loop, \tech appends both an example and a generation strategy to the input prompt, enabling the \generation \llm to continuously create new fuzzing inputs.

\begin{algorithm}[t!]
\setstretch{0.6}
\small
\caption{Fuzzing loop}

\label{alg:fuzzingloop}
\SetKwData{lmodel}{$\mathcal{M_{G}}$}
\SetKwFunction{system}{\sut}
\SetKwData{batchsize}{bs}
\SetKwData{inputprompt}{inputPrompt}
\SetKwData{timeout}{timeBudget}

\SetKwData{codesnippets}{fuzzingInputs}
\SetKwData{codeexample}{example}
\SetKwData{selectedinstruction}{instruction}
\SetKwData{genstrategy}{genStrats}
\SetKwData{instruct}{instruction}

\SetKwData{generationinstruction}{\generatenew}
\SetKwData{mutateinstruction}{\mutate}
\SetKwData{semanticinstruction}{\semanticequiv}
\SetKwData{}

\SetKwData{timeElapsed}{timeElapsed}

\SetKwData{potentialbugs}{bugs}

\SetKwProg{Fn}{Function}{:}{}
\SetKwFunction{oracle}{Oracle}

\SetKwFunction{FuzzingLoop}{FuzzingLoop}
\SetKwInOut{Input}{Input}
\SetKwInOut{Output}{Output}
\SetKw{Break}{break}

\Fn{\FuzzingLoop}{
    \Input{\inputprompt, \timeout}
    \Output{\potentialbugs}
    \BlankLine

    \genstrategy $\leftarrow$ [ \generationinstruction, \mutateinstruction, \semanticinstruction] \label{ts:genstrat}\\

    \codesnippets $\leftarrow$ \lmodel(\inputprompt + \generationinstruction) \label{ts:firstgen}\\

    \potentialbugs $\leftarrow$ \oracle(\codesnippets, \sut) \label{ts:oraclebugs} \\

    \While{\timeElapsed < \timeout}{\label{ts:loopstart}
        \codeexample $\leftarrow$ $\mathit{sample}$ (\codesnippets, \sut) \label{ts:selectcode}\\
        \selectedinstruction $\leftarrow$ $\mathit{sample}$ (\genstrategy) \label{ts:selectinstruct} \\
        \codesnippets $\leftarrow$ \lmodel(\inputprompt + \codeexample + \selectedinstruction) \label{ts:loopgen}\\
        \potentialbugs $\leftarrow$ \potentialbugs + \oracle(\codesnippets, \sut) \label{ts:oraclebugsloop} \\
    }\label{ts:loopend}
    \algorithmicreturn{ \potentialbugs }
}
\end{algorithm}

\subsubsection{Fuzzing Loop Algorithm}

\begin{figure}[t]
    \captionsetup{justification=centering}
    \centering
    \includegraphics[keepaspectratio=true,width=\linewidth]{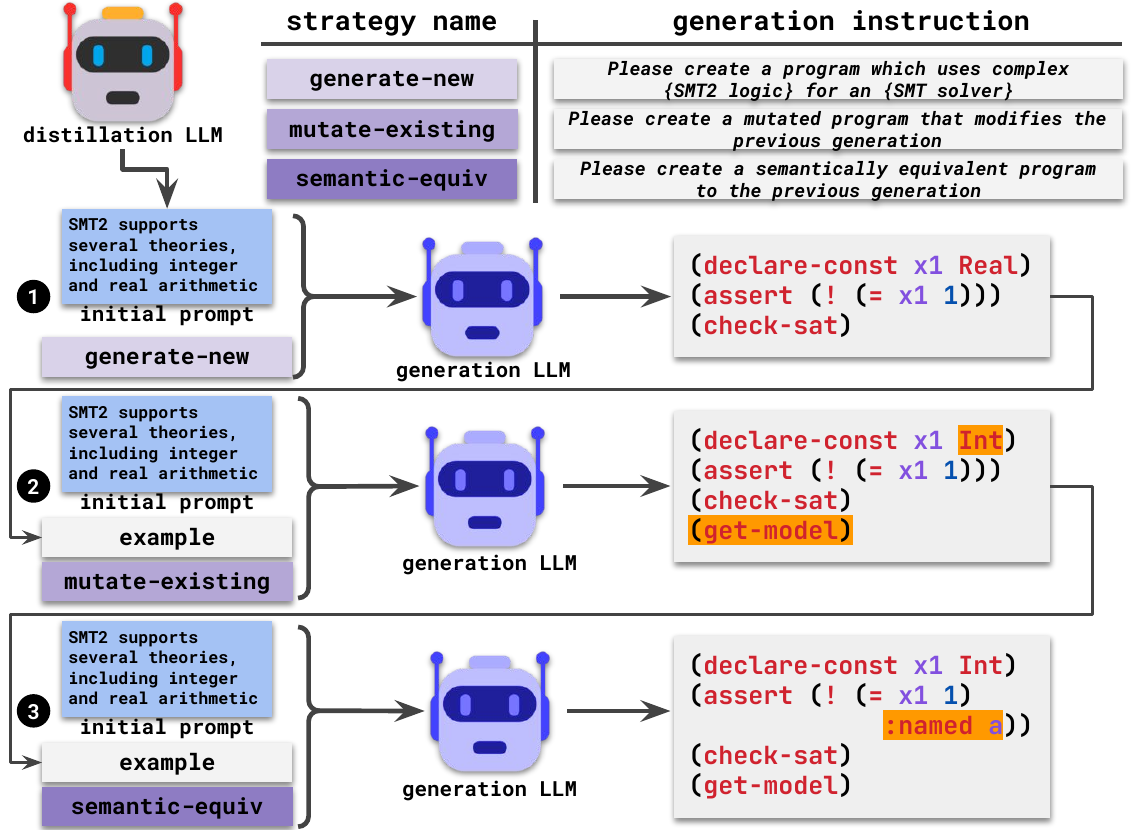}
    \caption{Fuzzing strategies and example of fuzzing loop.}
    \label{fig:genstrat}
\end{figure}

Algorithm~\ref{alg:fuzzingloop} describes the fuzzing loop. The inputs are the initial input prompt and the fuzzing budget.
The final output is a set of bugs identified by the user-defined oracle.
First, the algorithm initializes the generation strategies (\generatenew, \mutate, and \semanticequiv), which will be used to modify the input prompt during the fuzzing loop (line~\ref{ts:genstrat}).
Figure~\ref{fig:genstrat} (top-right) lists our three generation strategies along with their corresponding instructions.
For the first invocation of the \generation \llm, denoted with $\modelG$, the algorithm does not yet have any examples of fuzzing inputs.
Hence, it appends to the input prompt the \generatenew generation instruction, which guides the model toward producing a first batch of fuzzing inputs (line~\ref{ts:firstgen}).

Next, the algorithm enters the main fuzzing loop (lines~\ref{ts:loopstart}--\ref{ts:loopend}), which continuously updates the prompt to create new fuzzing inputs.
To this end, the algorithm selects an example from the previous batch of generated fuzzing inputs, randomly picking from all those fuzzing inputs that are valid for the \sut (line~\ref{ts:selectcode}).
In addition to the example, the algorithm also randomly picks one of the three generation strategies (line~\ref{ts:selectinstruct}).
The generation strategy either instructs the model to mutate the selected example (\mutate), to produce a fuzzing input that is semantically equivalent to the example (\semanticequiv), or to come up with a new fuzzing input (\generatenew).
The algorithm concatenates the initial input prompt, the selected example, and the selected generation strategy into a new prompt, and then queries the \generation \llm with this prompt to produce another batch of fuzzing inputs (line~\ref{ts:loopgen}).

The main fuzzing loop is repeated until the algorithm has exhausted the fuzzing budget.
For each created fuzzing input, \tech passes the input to the \sut.
If the user-defined oracle identifies an unexpected behavior, \eg a crash, then the algorithm adds a report to the set of detected bugs (lines~\ref{ts:oraclebugs} and~\ref{ts:oraclebugsloop}).

\subsubsection{Example: Fuzzing Loop}

Figure~\ref{fig:genstrat} illustrates how our fuzzing loop uses input examples and the generation strategies to create different fuzzing inputs.
In this case, we are fuzzing an SMT solver where the inputs are logic formulas written in the SMT2 language.
Initially \circled{1}, there are no examples, and hence, the algorithm uses the \generatenew strategy to synthesize new fuzzing inputs.
Next, taking a generated, valid fuzzing input as an example, the algorithm queries the model to create a new input \circled{2} based on the \mutate strategy, which aims to mutate the selected example.
We observe that the new fuzzing input subtly modifies the previous input by swapping the type of a variable as well as adding some computation.
In the next fuzzing iteration \circled{3}, the algorithm selects the previously generated fuzzing input as the example and uses the \semanticequiv generation strategy, which aims to create an input that does not modify the semantics of the given example.
This time, we observe that the new fuzzing input simply adds a syntax tag to the selected example.
In fact, the combination of generation strategies shown in the example helps \tech to generate a fuzzing input that causes an unexpected crash in the SMT solver.
The crash exposes one of the real-world bugs detected by \tech during our evaluation, which has been confirmed and fixed by developers.

\subsubsection{Oracle}

The fuzzing inputs produced by \tech during the fuzzing loop can be used to check the behavior of the \sut against an oracle to detect bugs.
The oracle is custom for each \sut, and it can be fully defined and customized by the user.
For example, when fuzzing C compilers, a user could define a differential testing oracle that compares the compiler behavior under different optimization levels~\cite{yang2011csmith}.
In this paper, we focus on simple and easy-to-define oracles, such as crashes due to segmentation faults and internal assertion failures, with more details discussed in Section~\ref{sec:target_sut}.

\section{Experimental Design}

We evaluate \tech on the following research questions:

\begin{itemize}[noitemsep, leftmargin=*, topsep=0pt]
    \item \textbf{RQ1:} How does \tech compare against existing fuzzers?
    \item \textbf{RQ2:} How effective is \tech in performing targeted fuzzing?
    \item \textbf{RQ3:} How do different components contribute to \tech's effectiveness?
    \item \textbf{RQ4:} What real-world bugs does \tech find?
\end{itemize}

\subsection{Implementation}

\tech is primarily implemented in Python.
The autoprompting and fuzzing loop components of \tech contain only 872 LoC.
Compared to traditional fuzzers, such as \csmith (>80K LoC), which need high manual effort to implement generators, \tech has a very lightweight implementation.
\tech uses \gpt{4}~\cite{openai2023gpt4} as the \distillation \llm to perform autoprompting since this model is the state-of-the-art for a wide range of NLP-based reasoning tasks~\cite{bubeck2023sparks}.
Specifically, we use the \CodeIn{gpt-4-0613} checkpoint with \CodeIn{max\_token} of 500 provided via the OpenAI API~\cite{gpt4endpoint}. 
\CodeIn{max\_token} forces the prompts to always fit within the context window of the \generation \llm. 
For autoprompting, we sample four candidate prompts, generate 30 fuzzing inputs each, and evaluate using a scoring function based on validity rate (as described in Section~\ref{sec:autoprompt_algorithm}). 
For the fuzzing loop, we use the Hugging Face implementation of the \starcoder~\cite{li2023starcoder} model as the \generation \llm, which is trained on over one trillion code tokens across over 80 languages.
Our default setting when generating fuzzing inputs uses a temperature of 1, a batch size of 30, a maximum output length of 1,024 using nucleus sampling~\cite{holtzman2019nucleus} with a top-p of 1.

\subsection{Systems Under Test and Baselines}
\label{sec:target_sut}

\begin{table}[t]\centering
\caption{\sut{s} and baseline tools.}
\label{tab:sutbaseline}
\scalebox{0.8}{
\begin{tabular}{lrrrr}\toprule
\textbf{Language} &\textbf{SUT(s)} &\textbf{Baseline tool(s)} &\textbf{Version} \\\midrule
C &GCC, Clang &\grayc~\cite{evenmendoza2023grayc}, \csmith~\cite{yang2011csmith} &GCC-13.1.1 \\
\cellcolor[HTML]{e1ecff}C++ &\cellcolor[HTML]{e1ecff}G++, Clang++ &\cellcolor[HTML]{e1ecff}\yarpgen~\cite{livinskii2020yarpgen} &\cellcolor[HTML]{e1ecff}G++-13.1.1 \\
SMT2 &Z3, CVC5 &\typefuzz~\cite{park2021generative} &CVC5-1.0.5 \\
\cellcolor[HTML]{e1ecff}Go &\cellcolor[HTML]{e1ecff}Go &\cellcolor[HTML]{e1ecff}\gofuzz~\cite{gofuzz} &\cellcolor[HTML]{e1ecff}go-1.20.6 \\
Java & javac & \hephaestus~\cite{chaliasos2022hephaestus} & OpenJDK-javac-18\\
\cellcolor[HTML]{e1ecff}Python &\cellcolor[HTML]{e1ecff}Qiskit &\cellcolor[HTML]{e1ecff}\morphq~\cite{paltenghiMorphQMetamorphicTesting2023} &\cellcolor[HTML]{e1ecff}qiskit-0.43.1 \\
\bottomrule
\end{tabular}
}
\end{table}

To demonstrate the generality of \tech, we evaluate it on six input languages and nine \sut{s}.
Table~\ref{tab:sutbaseline} shows each of the languages, \sut{s}, and the corresponding baseline tools.
Note that we compare coverage on one \sut per language, with the \sut versions used for coverage measurements shown in the last column of Table~\ref{tab:sutbaseline}.
Except for the coverage experiments, we perform fuzzing on the nightly release of each target.
Unless otherwise mentioned, we use unexpected compiler crashes as the oracle and consider a fuzzing input as valid if it compiles successfully.
Each baseline fuzzer is run with its default settings.
For baseline fuzzers that require input seeds, we use the default seed corpus provided in their replication repository. 
We now present more evaluation details for each \sut.

\subsubsection{C/C++ Compilers} We target the popular \gcc and \clang compilers and provide the standard C library documentation as user input to \tech by default. Our baselines include \csmith~\cite{yang2011csmith}, a classic generation-based C compiler fuzzer, and \grayc~\cite{evenmendoza2023grayc}{}, a recent mutation-based fuzzer that uses coverage feedback together with specialized mutation operators. 
For C++, we target new C++23 features by providing the C++23 standard documentation as input to \tech.
Our baseline is \yarpgen~\cite{livinskii2020yarpgen}, a generation-based fuzzer that extends \csmith with new language features in C++ and generation policies to trigger different compiler optimizations.

\subsubsection{SMT Solvers} We run \tech on \zee and \cvcf with commonly enabled developer settings, such as \CodeIn{debug} and \CodeIn{assertion}, following prior work~\cite{park2021generative, winterer-zhang-su-oopsla2020, winterer-zhang-su-pldi2020}.
\tech generates SMT formulas as fuzzing inputs using an overview documentation of the SMT2 language and SMT solver as input by default.
A fuzzing input is considered valid if the SMT solver returns either \CodeIn{SAT} or \CodeIn{UNSAT} without any error.
Our baseline is state-of-the-art \typefuzz~\cite{park2021generative}, which mutates existing SMT expressions based on newly generated expressions of the same type.

\subsubsection{Go Toolchain} We run \tech on the most recent version of Go.
By default, we use the Go standard library documentation as input to \tech.
As a baseline, we use \gofuzz~\cite{gofuzz}, a coverage-guided, mutation-based fuzzer designed for Go, which generates inputs for various Go standard libraries using handwritten templates.

\subsubsection{Java Compiler} We evaluate \tech on the OpenJDK Java compiler, javac, which compiles source code into bytecode.
Our default input is the latest standard Java API documentation page.
We compare against \hephaestus~\cite{chaliasos2022hephaestus}, a recent combined generation- and mutation-based fuzzer designed for JVM compilers and targeting type-related bugs.

\subsubsection{Quantum Computing Platform.}
We target \qiskit~\cite{QiskitQiskit2021}, a popular quantum computing framework~\cite{fingerhuthOpenSourceSoftware2018}.
\qiskit is built on top of Python, i.e., both the input program and the compilation are defined in Python code.
Thus, creating a valid input for \qiskit means using the \qiskit Python APIs in a meaningful way, \eg to create a quantum circuit.
It is challenging for traditional synthesis tools to handle dynamically typed general-purpose languages (like Python)~\cite{shi2022tf, gulwani2017program}, not to mention the additional API constraints and quantum-specific nature of many bugs~\cite{oopsla2022}, making fuzzing \qiskit a particularly difficult challenge.
Our baseline is \morphq~\cite{paltenghiMorphQMetamorphicTesting2023}, a recent fuzzer that uses a template- and grammar-based approach to generate valid quantum programs and then applies metamorphic transformations. 

Unlike for the other \sut{s}, which receive fuzzing inputs in a file, to invoke \qiskit, we must run the generated Python program itself.
As an oracle, we add statements at the end of the generated Python file, which collect all \CodeIn{QuantumCircuit} objects via Python's built-in introspection APIs and then apply two oracles on each circuit.
The two oracles are directly borrowed from previous work for a fair comparison~\cite{paltenghiMorphQMetamorphicTesting2023}.
The first oracle compiles the circuit via a \CodeIn{transpile} call with different optimization levels and reports any crash.
The second oracle converts the circuit to its lower-level QASM~\cite{crossOpenQuantumAssembly2017} representation and then reads it back, reporting any crash.

\subsection{Experimental Setup and Metrics}
\label{sec:evaluation_setup}

\parabf{Fuzzing campaigns.} For RQ1, we use a fuzzing budget of 24 hours (including autoprompting), which is used commonly in prior work~\cite{kleesEvaluatingFuzzTesting2018}. To account for variance, we repeat the experiment for both \tech and the baselines five times. 
Due to the high cost of experiments, for later RQs, we use a fuzzing budget of 10,000 generated fuzzing inputs and repeat four times for the ablation study.

\parabf{Environment.} Experiments are conducted on a 64-core workstation with 256 GB RAM running Ubuntu 20.04.5 LTS with 4 NVIDIA RTX A6000 GPUs (only one GPU is used per fuzzing run).

\parabf{Metrics.}
We use the widely adopted measure of \emph{code coverage} for evaluating fuzzing tools~\cite{bohme2022reliability, kleesEvaluatingFuzzTesting2018, wei2022free}. 
To be uniform, we report the line coverage for each of the targets studied in the evaluation. 
Following prior work~\cite{kleesEvaluatingFuzzTesting2018}, we use the Mann-Whitney U-test~\cite{mann1947test} to compute statistical significance and indicate significant (p < 0.05) coverage results in applicable tables (Tables~\ref{tab:gen_num_valid_rate} and~\ref{tab:ablation}) with *. 
We additionally measure the \emph{validity rate} (\% valid) of inputs as the percentage of fuzzing inputs generated that are valid and unique.
As \tech supports both general and targeted fuzzing, to assess the effectiveness of targeted fuzzing, we report the \emph{hit rate}, i.e., the percentage of fuzzing inputs that use a specific target feature (checked with simple regular expressions).
Finally, we also report the most important metric and goal of fuzzing: the number of bugs detected by \tech for each of our nine \sut{s}.

\section{Results}

\subsection{RQ1: Comparison against Existing Fuzzers}

\begin{figure*}[htb]
\captionsetup[subfigure]{aboveskip=-3pt,belowskip=-3pt}
\begin{subfigure}[b]{0.32\textwidth}
    \centering
    \includegraphics[width=\textwidth]{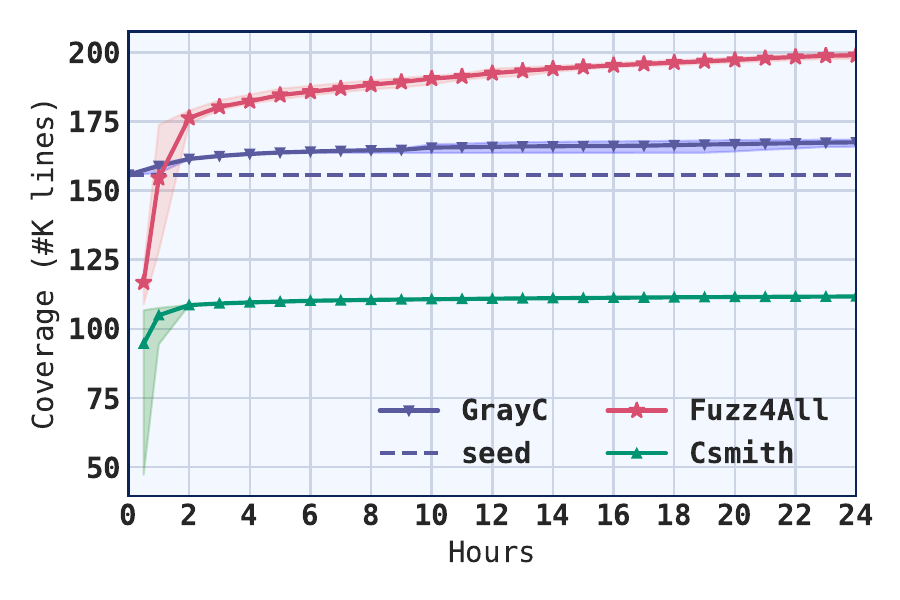}
    \caption{GCC}
    \label{fig:maincoverage-gcc}
\end{subfigure}
\begin{subfigure}[b]{0.32\textwidth}
    \centering
    \includegraphics[width=\textwidth]{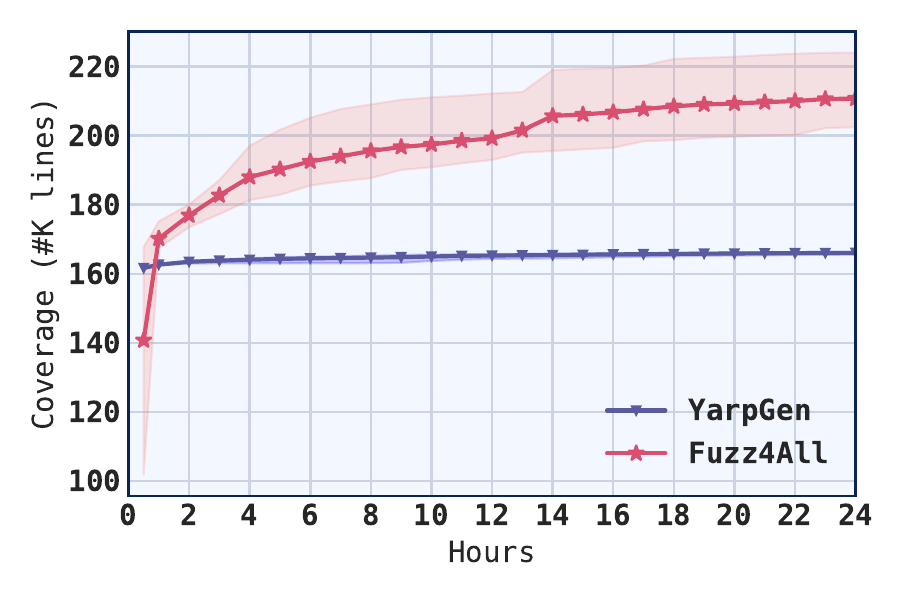}
    \caption{G++}
    \label{fig:maincoverage-g++}
\end{subfigure}
\begin{subfigure}[b]{0.32\textwidth}
    \centering
    \includegraphics[width=\textwidth]{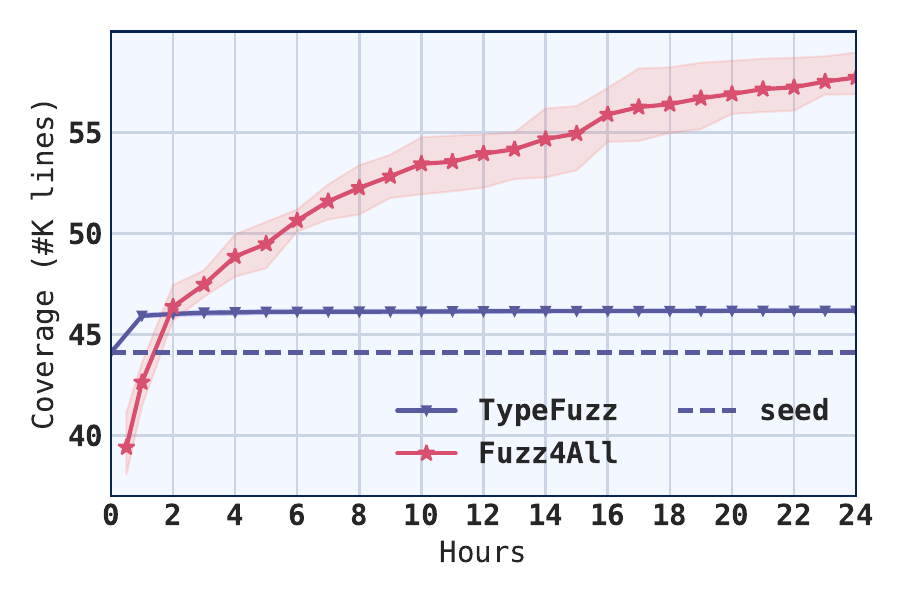}
    \caption{CVC5}
    \label{fig:maincoverage-cvc5}
\end{subfigure}
\begin{subfigure}[b]{0.32\textwidth}
    \centering
    \includegraphics[width=\textwidth]{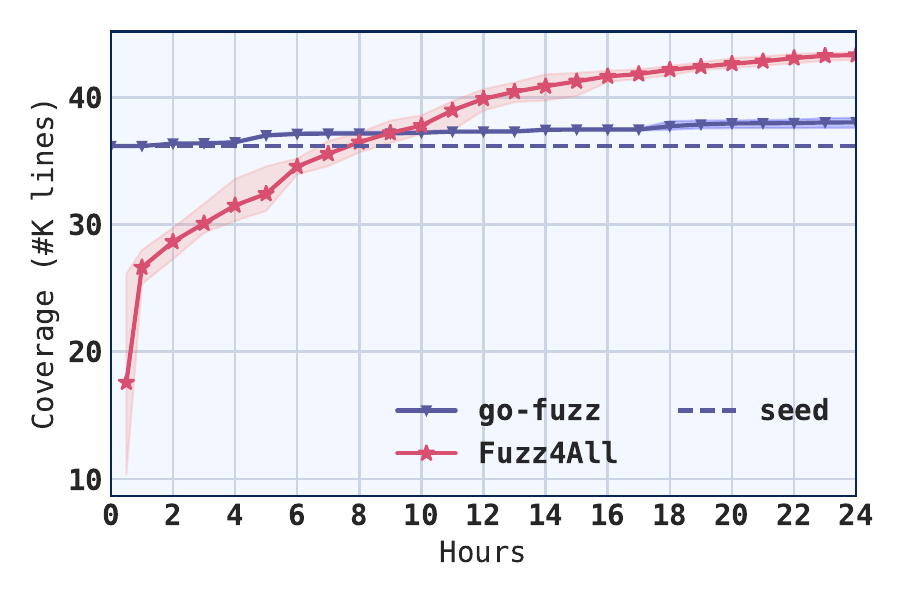}
    \caption{Go}
    \label{fig:maincoverage-go}
\end{subfigure}
\begin{subfigure}[b]{0.32\textwidth}
    \centering
    \includegraphics[width=\textwidth]{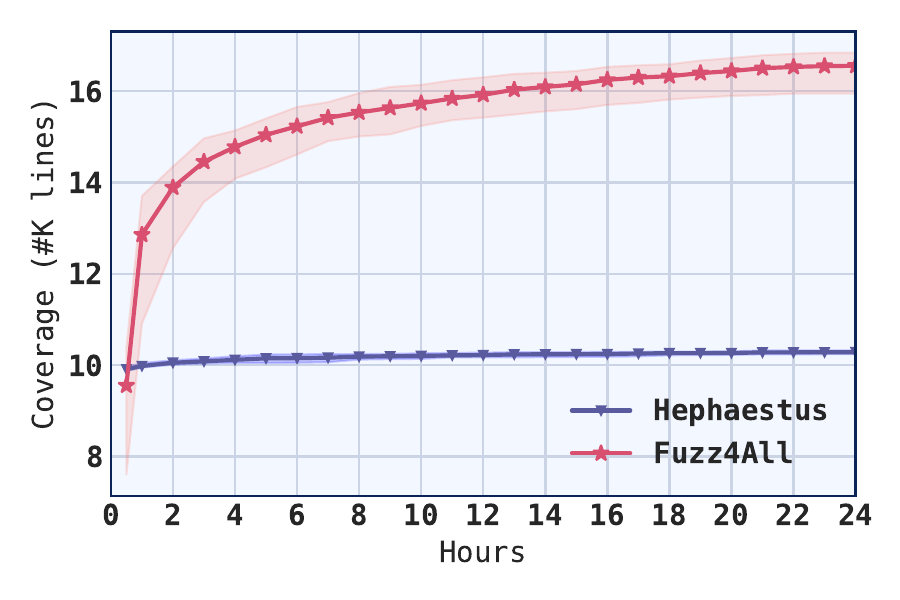}
    \caption{javac}
    \label{fig:maincoverage-javac}
\end{subfigure}
\begin{subfigure}[b]{0.32\textwidth}
    \centering
    \includegraphics[width=\textwidth]{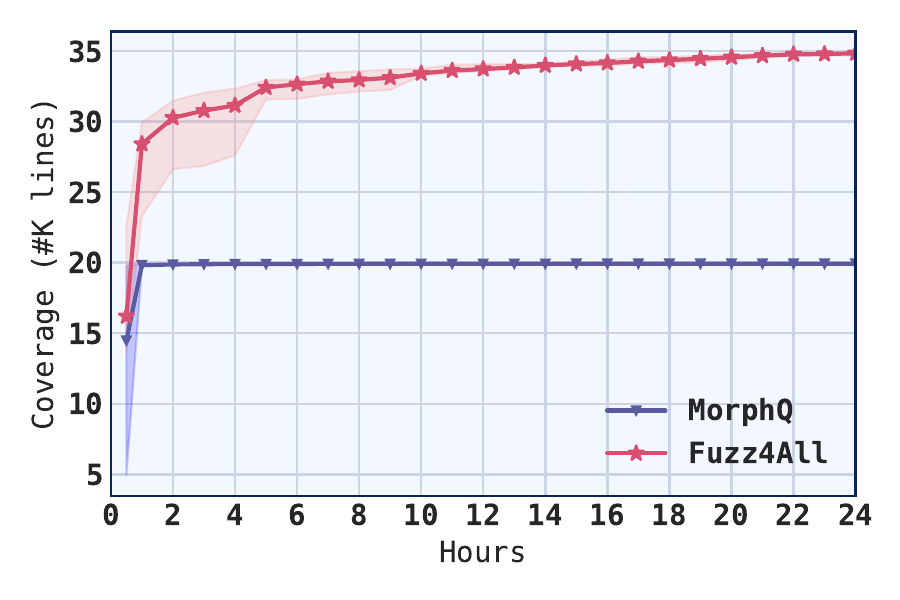}
    \caption{Qiskit}
    \label{fig:maincoverage-qiskit}
\end{subfigure}

\caption{Coverage trend of \tech against state-of-the-art fuzzers in a 24-hour fuzzing campaign.}

\label{fig:maincoverage}
\end{figure*}

\begin{table}[t]\centering
\caption{\tech against state-of-the-art fuzzers (* indicates statistically significant coverage improvement).}
\label{tab:gen_num_valid_rate}
\scalebox{0.85}{
\begin{tabular}{lrrrrrr}\toprule
\textbf{Target} &\textbf{Fuzzer} &\textbf{\# programs} &\textbf{ \% valid} &\multicolumn{2}{c}{\textbf{Coverage}} \\\midrule
\multirow{3}{*}{GCC} &\grayc &104,326 &95.96\% &167,453 & \\
&\csmith &61,883 &99.99\% &111,668 & \\
&\tech &44,324 &37.26\% &\textbf{*198,927} &\textbf{{\color[HTML]{009933}{+18.8\%}}} \\
\cellcolor[HTML]{e1ecff} &\cellcolor[HTML]{e1ecff}\yarpgen &\cellcolor[HTML]{e1ecff}255,581 &\cellcolor[HTML]{e1ecff}99.99\% &\cellcolor[HTML]{e1ecff}166,614 &\cellcolor[HTML]{e1ecff} \\
\multirow{-2}{*}{\cellcolor[HTML]{e1ecff}G++} &\cellcolor[HTML]{e1ecff}\tech &\cellcolor[HTML]{e1ecff}26,365 &\cellcolor[HTML]{e1ecff}40.74\% &\cellcolor[HTML]{e1ecff}\textbf{*210,743} &\cellcolor[HTML]{e1ecff}\textbf{{\color[HTML]{009933}{+26.5\%}}} \\
\multirow{2}{*}{CVC5} &\typefuzz &43,001 &93.24\% &46,174 & \\
&\tech &36,054 &47.63\% &\textbf{*57,674} &\textbf{{\color[HTML]{009933}{+24.9\%}}} \\
\cellcolor[HTML]{e1ecff} &\cellcolor[HTML]{e1ecff}\gofuzz &\cellcolor[HTML]{e1ecff}20,002 &\cellcolor[HTML]{e1ecff}100.00\% &\cellcolor[HTML]{e1ecff}38,024 &\cellcolor[HTML]{e1ecff} \\
\multirow{-2}{*}{\cellcolor[HTML]{e1ecff}Go} &\cellcolor[HTML]{e1ecff}\tech &\cellcolor[HTML]{e1ecff}22,817 &\cellcolor[HTML]{e1ecff}23.02\% &\cellcolor[HTML]{e1ecff}\textbf{*43,317} &\cellcolor[HTML]{e1ecff}\textbf{{\color[HTML]{009933}{+13.7\%}}} \\
\multirow{2}{*}{javac} &\hephaestus &728,217 &57.22\% &10,285 & \\
&\tech &31,967 &49.05\% &\textbf{*16,552} &\textbf{{\color[HTML]{009933}{+60.9\%}}} \\
\cellcolor[HTML]{e1ecff} &\cellcolor[HTML]{e1ecff}\morphq &\cellcolor[HTML]{e1ecff}38,474 &\cellcolor[HTML]{e1ecff}100.00\% &\cellcolor[HTML]{e1ecff}19,929 &\cellcolor[HTML]{e1ecff} \\
\multirow{-2}{*}{\cellcolor[HTML]{e1ecff}Qiskit} &\cellcolor[HTML]{e1ecff}\tech &\cellcolor[HTML]{e1ecff}33,454 &\cellcolor[HTML]{e1ecff}24.90\% &\cellcolor[HTML]{e1ecff}\textbf{*34,988} &\cellcolor[HTML]{e1ecff}\textbf{{\color[HTML]{009933}{+75.6\%}}} \\
\bottomrule
\end{tabular}
}
\end{table}

\subsubsection{Coverage over Time}

Figure~\ref{fig:maincoverage} shows the 24-hour coverage trend of \tech compared with the baselines, where the solid line shows average coverage and the area indicates the minimum and maximum across five runs. 
We observe that \tech achieves the highest coverage by the end of the fuzzing campaign across all targets, with an average improvement of \coverageimprove\% compared to the top performing baselines. 
Contrasting with generation-based fuzzers (\ie \yarpgen and \morphq), \tech is able to almost immediately achieve higher coverage, demonstrating the powerful generative ability of \llm{s} in producing diverse code snippets compared to traditional program generation techniques.
While mutation-based fuzzers (\ie{} \gofuzz and \grayc) are able to achieve higher coverage in the beginning through the use of high quality seeds, the coverage gained via mutations rapidly falls off and \tech is able to slowly but surely cover more code. 
Note that we include the autoprompting time as part of the fuzzing budget for a fair comparison, which incurs negligible overhead (avg. 2.3 minutes per fuzzing campaign). 

Unlike the baseline fuzzers, which reach a coverage plateau by the end of the 24-hour period, \tech keeps finding inputs that cover new code, even near the end of the fuzzing campaign.
Recall that during each iteration of \tech's fuzzing loop, the original input prompt is updated with both a new example and a generation strategy (Section~\ref{sec:fuzzing_loop}), nudging the \llm to generate new fuzzing inputs.
We hypothesize that this allows \tech to effectively generate new and diverse fuzzing inputs even after a long period of fuzzing, leading to sustained coverage increase.

\subsubsection{Generation Validity, Number, and Coverage}

We examine the number of fuzzing inputs generated and their validity rate across our studied \sut{s}.
In Table~\ref{tab:gen_num_valid_rate}, Column ``\# programs'' represents the number of unique inputs generated, ``\% valid'' is the percentage of fuzzing inputs that are valid, and ``Coverage'' shows the final coverage obtained by each fuzzer along with the relative improvement over the best baseline. 
We first observe that almost all traditional fuzzing tools can achieve a very high validity rate apart from \hephaestus, which purposefully generates invalid code (focused on incorrect types) to check for miscompilation bugs.
In contrast, \tech has a lower percentage of valid fuzzing inputs generated (56.0\% average reduction compared to baseline tools).
Furthermore, the raw number of fuzzing inputs generated by baseline tools are also much higher. 
By using an \llm as the generation engine, \tech is bottlenecked by GPU inference, leading to 43.0\% fewer fuzzing inputs compared to traditional fuzzers.

In spite of the lower validity rate and number of fuzzing inputs, \tech generates much more diverse programs compared to traditional fuzzing tools, as evidenced by the high coverage obtained (+\coverageimprove\% average increase). 
Additionally, even invalid code snippets that are close to valid can be useful for fuzzing, as they allow for finding bugs in the validation logic of the \sut.
In Section~\ref{sec:bugs}, we further describe the various types of bugs detected by \tech, with both valid and invalid code snippets, to additionally showcase the benefit of generating diverse fuzzing inputs.

We note that \tech achieves a wide range of validity rates and numbers of fuzzing inputs across different \sut{s}. 
The number of fuzzing inputs varies across targets due to the varying cost to invoke the \sut after each fuzzing iteration for bug detection. 
Regarding validity rate, a general-purpose programming language, such as C, has a relatively lower validity rate compared to domain-specific languages, such as the SMT2 language used for SMT solvers.
A more rigorous language, \eg Go, which does not allow any declared but unused variables, has an even lower validity rate.
We also observe a low validity rate for fuzzing quantum computing platforms.
As quantum computing is an emerging area with its own set of library APIs, the \generation \llm may not have seen as many examples of quantum programs during its training as for more established languages.
Nevertheless, \tech is still able to leverage user-provided documentation to generate interesting fuzzing inputs that use quantum library APIs and achieve an impressive coverage improvement (+75.6\%) compared to the state-of-the-art fuzzer.

\subsection{RQ2: Effectiveness of Targeted Fuzzing}

\begin{table}[t]\centering
\caption{Hit rate and coverage during targeted fuzzing.}\label{tab:target_fuzzing}
\scalebox{0.8}{
\begin{tabular}{llrrrrr}\toprule

&\multicolumn{5}{c}{\textbf{C targeted campaign (keywords) }} \\
& &\CodeIn{typedef} &\CodeIn{union} &\CodeIn{goto} &General \\\midrule
\multirow{3}{*}{\rotatebox[origin=c]{90}{Hit rate}} &\CodeIn{typedef} &\cellcolor[HTML]{e1ecff}\textbf{83.11\%} &47.16\% &0.48\% &4.38\% \\
&\CodeIn{union} &10.80\% &\cellcolor[HTML]{e1ecff}\textbf{80.43\%} &0.10\% &0.32\% \\
&\CodeIn{goto} &0.22\% &0.11\% &\cellcolor[HTML]{e1ecff}\textbf{77.62\%} &1.16\% \\ \midrule
\multicolumn{2}{l}{Coverage} &123,226 &125,041 &120,452 &188,148 \\ \midrule
&\multicolumn{5}{c}{\textbf{C++ targeted campaign (built-in functions)} } \\
& &\CodeIn{apply} &\CodeIn{expected} &\CodeIn{variant} &General \\ \midrule
\multirow{3}{*}{\rotatebox[origin=c]{90}{Hit rate}} &\CodeIn{apply} &\cellcolor[HTML]{e1ecff}\textbf{70.23\%} &0.41\% &0.68\% &0.32\% \\
&\CodeIn{expected} &0.26\% &\cellcolor[HTML]{e1ecff}\textbf{79.72\%} &0.94\% &1.33\% \\
&\CodeIn{variant} &1.16\% &5.98\% &\cellcolor[HTML]{e1ecff}\textbf{93.19\%} &3.63\% \\ \midrule
\multicolumn{2}{l}{Coverage} &182,261 &175,963 &182,333 &193,254 \\\midrule
&\multicolumn{5}{c}{\textbf{SMT targeted campaign (theories)} } \\
& &\CodeIn{Array} &\CodeIn{BitVec} &\CodeIn{Real} &General \\ \midrule
\multirow{3}{*}{\rotatebox[origin=c]{90}{Hit rate}} &\CodeIn{Array} &\cellcolor[HTML]{e1ecff}\textbf{82.23\%} &2.08\% &1.44\% &11.07\% \\
&\CodeIn{BitVec} &2.57\% &\cellcolor[HTML]{e1ecff}\textbf{88.48\%} &0.86\% &5.46\% \\
&\CodeIn{Real} &1.45\% &0.17\% &\cellcolor[HTML]{e1ecff}\textbf{96.01\%} &17.36\% \\ \midrule
\multicolumn{2}{l}{Coverage} &46,392 &48,841 &47,619 &52,449 \\ \midrule
&\multicolumn{5}{c}{\textbf{Go targeted campaign (built-in libraries)}} \\
& &\CodeIn{atomic} &\CodeIn{atomic} &\CodeIn{heap} &General \\ \midrule
\multirow{3}{*}{\rotatebox[origin=c]{90}{Hit rate}} &\CodeIn{atomic} &\cellcolor[HTML]{e1ecff}\textbf{90.09\%} &0.04\% &0.06\% &1.01\% \\
&\CodeIn{big} &0.18\% &\cellcolor[HTML]{e1ecff}\textbf{97.20\%} &0.23\% &3.63\% \\
&\CodeIn{heap} &0.30\% &0.04\% &\cellcolor[HTML]{e1ecff}\textbf{91.18\%} &2.22\% \\ \midrule
\multicolumn{2}{l}{Coverage} &10,156 &12,986 &9,790 &37,561 \\ \midrule
&\multicolumn{5}{c}{\textbf{Java targeted campaign (keywords)}} \\
& &\CodeIn{instanceof} &\CodeIn{synchronized} &\CodeIn{finally} &General \\ \midrule
\multirow{3}{*}{\rotatebox[origin=c]{90}{Hit rate}} &\CodeIn{instanceof} &\cellcolor[HTML]{e1ecff}\textbf{88.00\%} &0.08\% &0.85\% &1.86\% \\
&\CodeIn{synchronized} &0.16\% &\cellcolor[HTML]{e1ecff}\textbf{94.80\%} &0.16\% &0.85\% \\
&\CodeIn{finally} &0.51\% &3.17\% &\cellcolor[HTML]{e1ecff}\textbf{78.62\%} &0.82\% \\ \midrule
\multicolumn{2}{l}{Coverage} &14,546 &13,972 &13,203 &16,128 \\ \midrule
&\multicolumn{5}{c}{\textbf{Qiskit targeted campaign (APIs)}} \\
& &\CodeIn{switch} &\CodeIn{for loop} &\CodeIn{linear} &General \\ \midrule
\multirow{3}{*}{\rotatebox[origin=c]{90}{Hit rate}} &\CodeIn{switch} &\cellcolor[HTML]{e1ecff}\textbf{71.76\%} &0.00\% &0.00\% &0.00\% \\
&\CodeIn{for loop} &0.17\% &\cellcolor[HTML]{e1ecff}\textbf{75.97\%} &0.00\% &0.00\% \\
&\CodeIn{linear} &0.00\% &0.00\% &\cellcolor[HTML]{e1ecff}\textbf{54.79\%} &0.00\% \\ \midrule
\multicolumn{2}{l}{Coverage} &30,597 &26,703 &29,535 &33,853 \\
\bottomrule
\end{tabular}
}
\end{table}

We now evaluate the ability of \tech to perform targeted fuzzing, i.e., to generate fuzzing inputs that focus on a particular feature.
For each target \sut and language, we target three different example features and compare them to the setup with general user input, as used for RQ1 (described in Section~\ref{sec:evaluation_setup}).
These features are built-in libraries or functions/APIs (Go, C++ and Qiskit), language keywords (C and Java), and theories (SMT).
The user input for the targeted fuzzing runs is documentation of the particular feature we are focusing on. Table~\ref{tab:target_fuzzing} shows the results of targeted fuzzing as well as the default general fuzzing used in RQ1.
Each column represents a targeted fuzzing run where we focus on one feature.
The value in each cell shows the hit rate of the feature (Section~\ref{sec:evaluation_setup}) for a particular fuzzing run.
We also include the coverage results obtained.

We observe that targeting a specific feature yields a high amount of fuzzing inputs that directly use the feature, with an average hit rate of 83.0\%. 
This result demonstrates that \tech indeed performs targeted fuzzing by prompting the \generation \llm with an input prompt that describes a particular feature. 
Furthermore, we observe that fuzzing on features that are related can lead to a moderately high cross-feature hit rate (\ie hit rate of feature \CodeIn{X} on fuzzing run for feature \CodeIn{Y}). 
For example, the C keywords \CodeIn{typedef} and \CodeIn{union} are both related to type operations, and hence, their cross-feature hit rate is high compared to an unrelated feature, such as \CodeIn{goto}.
As shown in Table~\ref{tab:target_fuzzing}, a general fuzzing approach, while achieving the highest overall code coverage, can be extremely inefficient in targeting a specific feature (average 96.0\% reduction in hit rate compared with \tech's targeted fuzzing).
For example, in \qiskit, the general fuzzing campaign has a 0\% hit rate of the three target features.
This can be explained by the fact that these features were added recently to \qiskit and are not yet widely used, thus being extremely rare in the \llm training data.
However, by providing suitable user input during the targeted fuzzing campaign, \tech can successfully generate fuzzing inputs that use these new features. 
This ability of \tech will be valuable to developers who want to test novel features or components of a \sut.

\subsection{RQ3: Ablation Study}
\label{sec:ablation_study}

\begin{table*}[!htp]\centering
\caption{Effectiveness of variants (* indicates statistically significant coverage improvement compared w/ 2nd best variant).}

\label{tab:ablation}

\scalebox{0.76}{
\begin{tabular}{lllrrrrrrrrrrrrr}\toprule
\multirow{2}{*}{} &\multirow{2}{*}{\textbf{Variants}} &\multirow{2}{*}{\textbf{Description}} &\multicolumn{2}{c}{\textbf{C}} &\multicolumn{2}{c}{\cellcolor[HTML]{e1ecff}\textbf{C++}} &\multicolumn{2}{c}{\textbf{SMT}} &\multicolumn{2}{c}{\cellcolor[HTML]{e1ecff}\textbf{Go}} &\multicolumn{2}{c}{\textbf{Java}} &\multicolumn{2}{c}{\cellcolor[HTML]{e1ecff}\textbf{Qiskit}} \\\cmidrule{4-15}
& & &Cov. &\% valid &\cellcolor[HTML]{e1ecff}Cov. &\cellcolor[HTML]{e1ecff}\% valid &Cov. &\% valid &\cellcolor[HTML]{e1ecff}Cov. &\cellcolor[HTML]{e1ecff}\% valid &Cov. &\% valid &\cellcolor[HTML]{e1ecff}Cov. &\cellcolor[HTML]{e1ecff}\% valid \\\midrule
\multirow{3}{*}{\rotatebox[origin=c]{90}{\makecell{Auto\\prompt.}}} &\noinput &no initial prompt &127,261 &42.57\% &\cellcolor[HTML]{e1ecff}181,493 &\cellcolor[HTML]{e1ecff}51.63\% &50,838 &49.49\% &\cellcolor[HTML]{e1ecff}35,765 &\cellcolor[HTML]{e1ecff}39.54\% &14,374 &50.25\% &\cellcolor[HTML]{e1ecff}31,701 &\cellcolor[HTML]{e1ecff}34.63\% \\
&\documentation &use user-provided input &137,204 &33.95\% &\cellcolor[HTML]{e1ecff}189,030 &\cellcolor[HTML]{e1ecff}33.79\% &49,697 &39.49\% &\cellcolor[HTML]{e1ecff}36,168 &\cellcolor[HTML]{e1ecff}16.84\% &15,445 &37.64\% &\cellcolor[HTML]{e1ecff}31,922 &\cellcolor[HTML]{e1ecff}22.74\% \\
&\autoprompt &apply autoprompting &182,530 &39.09\% &\cellcolor[HTML]{e1ecff}190,318 &\cellcolor[HTML]{e1ecff}36.62\% &51,496 &45.04\% &\cellcolor[HTML]{e1ecff}36,732 &\cellcolor[HTML]{e1ecff}24.87\% &15,838 &45.54\% &\cellcolor[HTML]{e1ecff}32,691 &\cellcolor[HTML]{e1ecff}29.12\% \\ \midrule
\multirow{3}{*}{\rotatebox[origin=c]{90}{\makecell{Fuzzing\\loop}} } &\noloop &\generatenew w/o example &143,349 &34.23\% &\cellcolor[HTML]{e1ecff}190,288 &\cellcolor[HTML]{e1ecff}28.25\% &50,089 &18.41\% &\cellcolor[HTML]{e1ecff}35,839 &\cellcolor[HTML]{e1ecff}19.38\% &15,444 &44.69\% &\cellcolor[HTML]{e1ecff}32,663 &\cellcolor[HTML]{e1ecff}24.04\% \\
&\addexample &\generatenew w/ example &182,530 &39.09\% &\cellcolor[HTML]{e1ecff}190,318 &\cellcolor[HTML]{e1ecff}36.62\% &51,496 &45.04\% &\cellcolor[HTML]{e1ecff}36,732 &\cellcolor[HTML]{e1ecff}24.87\% &15,838 &45.54\% &\cellcolor[HTML]{e1ecff}32,691 &\cellcolor[HTML]{e1ecff}29.12\% \\
&\tech &\all strategies w/ example &\textbf{185,491} &40.58\% &\cellcolor[HTML]{e1ecff}\textbf{*193,845} &\cellcolor[HTML]{e1ecff}41.22\% &\textbf{*53,069} &50.06\% &\cellcolor[HTML]{e1ecff}\textbf{*37,981} &\cellcolor[HTML]{e1ecff}32.00\% &\textbf{*16,209} &50.99\% &\cellcolor[HTML]{e1ecff}\textbf{*33,913} &\cellcolor[HTML]{e1ecff}27.45\% \\
\bottomrule
\end{tabular}
}
\end{table*}

To study how each component of \tech contributes to the overall fuzzing effectiveness, we conduct an ablation study based on the two key components of \tech:
(a) Autoprompting, the type of initial input prompt provided to the \generation \llm; 
(b) Fuzzing loop, the use of selected examples and generation strategies.
We study three variants for each of the two key components. 
Table~\ref{tab:ablation} shows the coverage and validity rate of our studied variants.

\subsubsection{Autoprompting}
First, we examine the effect of different initial inputs provided to the \generation \llm. To reduce the impact of additional factors, we fix the generation strategy to only use \generatenew and study three variants\footnote{The impact of additional generation strategies can be found in Section~\ref{sec:fl-study}.}:  
1) \textit{\noinput} does not use any initial prompts%
2) \textit{\documentation} directly uses the raw user input as the initial prompt,
3) \textit{\autoprompt} applies autoprompting to generate the initial prompt.
We observe that across all studied languages, the \noinput variant achieves the lowest coverage.
In \noinput, we do not provide any initial prompt, which provides useful information on the features we want to generate fuzzing inputs for.
As such, the \llm can only generate simple code snippets with high validity rate but is less effective in covering the \sut.
We observe a coverage boost as we use the \documentation variant, where we provide the raw documentation as the initial prompt.
However, we can further improve both the code coverage and the validity rate by using our autoprompting stage to distill the user input into a concise but informative prompt (\autoprompt), instead of using the raw user input.
Directly using the user-provided input may include information that is irrelevant for fuzzing, leading to both a lower validity rate (as the \generation \llm may struggle to understand the raw documentation) and lower coverage (since, unlike our autoprompting generated prompt, the raw documentation is not designed to be used for \llm generation).

\subsubsection{Fuzzing loop.}
\label{sec:fl-study}
Next, we examine the different variants of our fuzzing loop setup by keeping the initial prompt the same (by using the default autoprompting): 
1) \textit{\noloop} does not select an example during the fuzzing loop (\ie it continuously samples from the same initial prompt),
2) \textit{\addexample} selects an example but only uses the \generatenew instruction\footnote{Note that \smallautoprompt and \smalladdexample are the same variant, but we include them separately for ease of comparison.},
3) \tech{} is the full approach with all generation strategies used. 
We first observe that by only sampling from the same input (\noloop), \llm{s} will often repeatedly generate the same or similar fuzzing inputs. 
On average, 8.0\% of the fuzzing inputs generated are repeated in \noloop compared to only 4.7\% when using the full \tech approach. 
Adding an example to the input prompt (\addexample) avoids sampling from the same distribution and improves both the coverage and the validity rate. 
Finally, the full \tech approach achieves the highest coverage across all \sut{s}.
Compared to the \addexample variant (the second-best), the full \tech adds additional generation strategies, \semanticequiv and \mutate, which provide useful instructions to the \generation \llm.

\subsection{RQ4: Bug Finding}
\label{sec:bugs}

\begin{table}[t]\centering
\caption{Summary of \tech-detected bugs.}\label{tab:bugs}
\scalebox{0.85}{
\begin{tabular}{lrrrrrr}\toprule
&\multirow{2}{*}{\textbf{Total}} &\multicolumn{2}{c}{\textbf{Confirmed}} &\multirow{2}{*}{\textbf{Pending}} &\multirow{2}{*}{\textbf{Won't fix}} \\\cmidrule{3-4}
& &\textbf{Unknown} &\textbf{Known} & & \\\midrule
GCC &30 &14 &11 &5 &0 \\
Clang &27 &18 &9 &0 &0 \\
CVC5 &9 &7 &2 &0 &0 \\
Z3 &14 &12 &0 &0 &2 \\
Go &4 &2 &2 &0 &0 \\
Java &3 &3 &0 &0 &0 \\
Qiskit &11 &8 &2 &1 &0 \\ \midrule
Total &98 &64 &26 &6 &2 \\
\bottomrule
\end{tabular}
}
\end{table}

\begin{figure}[t]
    \captionsetup[subfigure]{aboveskip=-.5pt,justification=justified,singlelinecheck=false}
    \centering

    \begin{subfigure}[b]{\linewidth}
        \centering
        \includegraphics[width=\linewidth]{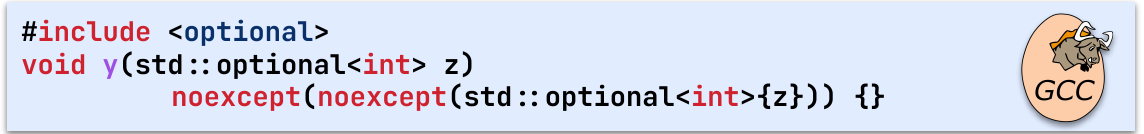}
        \caption{GCC bug: Internal compiler error (segmentation fault)}
        \label{fig:bug_gcc}
    \end{subfigure}
    \begin{subfigure}[b]{\linewidth}
        \centering
        \includegraphics[width=\linewidth]{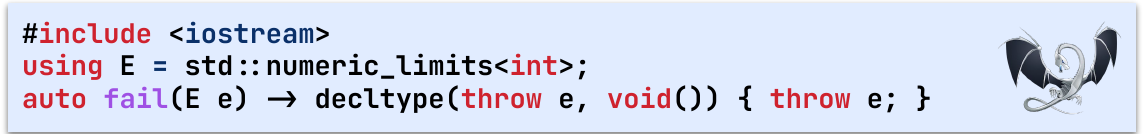}
        \caption{Clang bug: Segmentation fault}
        \label{fig:bug_clang}
    \end{subfigure}
    \begin{subfigure}[b]{\linewidth}
        \centering
        \includegraphics[width=\linewidth]{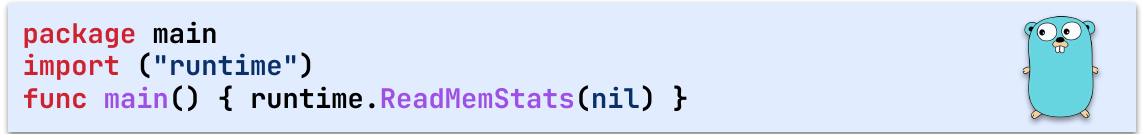}
        \caption{Go bug: Segmentation violation}
        \label{fig:bug_go}
    \end{subfigure}
    \begin{subfigure}[b]{\linewidth}
        \centering
        \includegraphics[width=\linewidth]{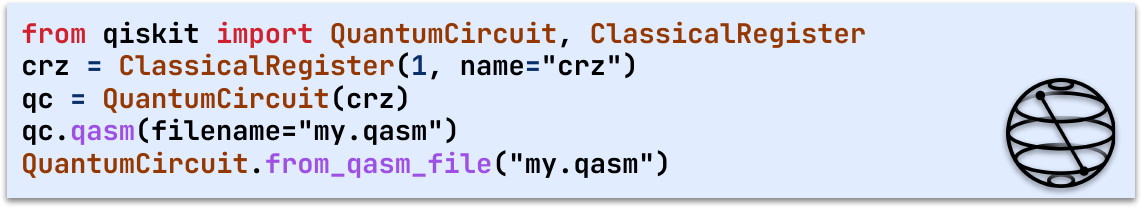}
        \caption{Qiskit bug: Crash}
        \label{fig:bug_qiskit}
    \end{subfigure}
    
    \caption{Exemplary bugs found by \tech.}
    \label{fig:bugs}
\end{figure}

Table~\ref{tab:bugs} summarizes the bugs found by \tech on our nine studied \sut{s}.
In total, \tech detects \totalbugs bugs, with \totalconfirmedbugs bugs already confirmed by the developers as previously unknown. 
These results not only demonstrate the practical effectiveness of \tech in finding large amounts of bugs but also the promised generality of \tech across languages and \sut{s}.
A detailed list of reported bugs and issue links can be found in our artifact.

\subsubsection{Examples}

Figure~\ref{fig:bug_gcc} shows a bug found in \gcc when using
\CodeIn{noexcept(x)}, a C++ feature that specifies a function is non-throwing if \CodeIn{x} evaluates to \CodeIn{true}. In this example bug, \tech generates a rather complex code using \CodeIn{std::optional}, which indicates that a particular value may or may not be present at runtime.
While this code is valid and should compile correctly, this combination of difficult runtime dependencies cause GCC to crash with an \CodeIn{internal compiler error}.
We note that this bug cannot be found by prior techniques since they simply do not support the \CodeIn{noexcept} feature.
The developers have already confirmed and fixed this bug. Interestingly, they even added a slightly modified version of our submitted code snippet to the official test suite of GCC.

Figure~\ref{fig:bug_clang} shows a bug found in \clang, where the invalid code leads to a segmentation fault. \tech uses an unusual syntax for function declaration (\ie{} \CodeIn{auto x (...) -> return\_type} ), which makes use of the \CodeIn{decltype} operation in C++. However, the bug occurs when the \CodeIn{throw} statement inside of the \CodeIn{decltype} is evaluated first, skipping the evaluation of the return type since \CodeIn{throw} exits the scope early and crashes \clang.
This code, while invalid, is still useful to reveal a bug in the \clang frontend as confirmed by the developers.
Additionally, prior fuzzing tools can hardly find this bug since they typically focus on generating valid code only and do not handle the especially difficult-to-model \CodeIn{decltype} function.

Figure~\ref{fig:bug_go} shows a bug found in Go where a \CodeIn{nil} input causes a segmentation fault instead of producing a useful failure message.
This bug is found by targeting the \CodeIn{runtime} Go standard library, where we provide the documentation, which includes the description of the \CodeIn{ReadMemStats} function.
The bug has been confirmed and fixed by the developers.
While this bug might look simple (invoking a singular function), it cannot be found by the \gofuzz baseline simply because \gofuzz requires manually written templates to target specific libraries, and \CodeIn{runtime} is not a part of any such template.
With \tech, users can directly target any Go standard libraries by providing relevant input information (\eg documentation).

Figure~\ref{fig:bug_qiskit} shows a bug found in \qiskit's QASM exporter.
A quantum program, represented by the \CodeIn{qc} variable, is exported to QASM, a low level representation, silently generating an invalid output file, which leads to a crash when being reimported.
The problem is that the exporter represents the register in QASM using its name as identifier, \ie \CodeIn{"crz"}, which also is the name of a well-known operation of the QASM language, thus making the generated code ambiguous.
Note that prior work~\cite{paltenghiMorphQMetamorphicTesting2023} could not find this bug because they use pre-defined templates with only anonymous registers, whereas \tech effectively leverages the quantum knowledge of \llm{s} to inject a meaningful string literal for detecting this bug.

\section{Threats to Validity}

\parabf{Internal.} The main internal threat comes from the implementation of \tech. To address this, we performed code reviews and testing to ensure correctness. Furthermore, we run each baseline from their provided replication package whenever possible.

\parabf{External.} The main external threat is our evaluation targets. To support our generality claim, we apply \tech on nine different \sut{s} across six languages. Additionally, to account for variance in long fuzzing runs, we repeat the 24-hour fuzzing campaign five times and check for statistically significant results. 
Since the \generation \llm leverages the knowledge acquired during its training done within the last year, reapplying \tech using the exact checkpoint of the \llm (\starcoder) used in this work might degrade the effectiveness in the future due to data-shift.
\tech can mitigate this using the autoprompting step where more up-to-date documentation/example code allows the model to also generate up-to-date fuzzing inputs. 
One additional threat comes from the use of the \distillation \llm to generate the initial inputs, where the \llm may ``hallucinate'', i.e., produce made-up or inaccurate information~\cite{guo2022survey}
. 
This limitation is common to most pipelines that use \llm{s}, and we hope to address it in our future work.

\section{Conclusion}

We present \tech, a universal fuzzer leveraging \llm{s} to support both general and targeted fuzzing of arbitrary \sut{s} that take in a multitude of programming languages. \tech uses a novel autoprompting stage to produce input prompts that concisely summarize the user-provided inputs. In its fuzzing loop, \tech iteratively updates the initial input prompt with both code examples and generation strategies aimed at producing diverse fuzzing inputs. Evaluation results on nine different \sut{s} across six different languages demonstrate that \tech is able to significantly improve coverage compared to state-of-the-art tools. 
Furthermore, \tech is able to detect \totalbugs bugs with \totalconfirmedbugs already confirmed by developers as previously unknown.

\section*{Data Availability}

Our code and data are available at: \url{https://doi.org/10.5281/zenodo.10456883} and \url{https://github.com/fuzz4all/fuzz4all}
\section*{Acknowledgment}
This work was supported by the National Science Foundation (grants CCF-2131943 and CCF-2141474), Kwai Inc., the European Research Council (ERC, grant agreement 851895), and the German Research Foundation within the ConcSys and DeMoCo projects.

\bibliographystyle{ACM-Reference-Format}
\bibliography{references, phd-mattepalte, referencesMichael}


\begin{thebibliography}{88}

%%% the Web address.

\ifx \showCODEN    \undefined \def \showCODEN     #1{\unskip}     \fi
\ifx \showDOI      \undefined \def \showDOI       #1{#1}\fi
\ifx \showISBNx    \undefined \def \showISBNx     #1{\unskip}     \fi
\ifx \showISBNxiii \undefined \def \showISBNxiii  #1{\unskip}     \fi
\ifx \showISSN     \undefined \def \showISSN      #1{\unskip}     \fi
\ifx \showLCCN     \undefined \def \showLCCN      #1{\unskip}     \fi
\ifx \shownote     \undefined \def \shownote      #1{#1}          \fi
\ifx \showarticletitle \undefined \def \showarticletitle #1{#1}   \fi
\ifx \showURL      \undefined \def \showURL       {\relax}        \fi
% invisible to TeX
\providecommand\bibfield[2]{#2}
\providecommand\bibinfo[2]{#2}
\providecommand\natexlab[1]{#1}
\providecommand\showeprint[2][]{arXiv:#2}

\bibitem[Qis(2021)]%
        {QiskitQiskit2021}
 \bibinfo{year}{2021}\natexlab{}.
\newblock \bibinfo{title}{Qiskit/Qiskit}.
\newblock \bibinfo{howpublished}{https://github.com/Qiskit/qiskit}.
\newblock


\bibitem[cpp(2023)]%
        {cppexpecteddocumentation}
 \bibinfo{year}{2023}\natexlab{}.
\newblock \bibinfo{title}{{std::expected}}.
\newblock
  \bibinfo{howpublished}{\url{https://en.cppreference.com/w/cpp/utility/expected}}.
\newblock


\bibitem[Aschermann et~al\mbox{.}(2019)]%
        {aschermann2019nautilus}
\bibfield{author}{\bibinfo{person}{Cornelius Aschermann},
  \bibinfo{person}{Tommaso Frassetto}, \bibinfo{person}{Thorsten Holz},
  \bibinfo{person}{Patrick Jauernig}, \bibinfo{person}{Ahmad-Reza Sadeghi},
  {and} \bibinfo{person}{Daniel Teuchert}.} \bibinfo{year}{2019}\natexlab{}.
\newblock \showarticletitle{NAUTILUS: Fishing for Deep Bugs with Grammars.}. In
  \bibinfo{booktitle}{\emph{NDSS}}.
\newblock


\bibitem[Bang et~al\mbox{.}(2023)]%
        {bang2023multitask}
\bibfield{author}{\bibinfo{person}{Yejin Bang}, \bibinfo{person}{Samuel
  Cahyawijaya}, \bibinfo{person}{Nayeon Lee}, \bibinfo{person}{Wenliang Dai},
  \bibinfo{person}{Dan Su}, \bibinfo{person}{Bryan Wilie},
  \bibinfo{person}{Holy Lovenia}, \bibinfo{person}{Ziwei Ji},
  \bibinfo{person}{Tiezheng Yu}, \bibinfo{person}{Willy Chung},
  {et~al\mbox{.}}} \bibinfo{year}{2023}\natexlab{}.
\newblock \showarticletitle{A multitask, multilingual, multimodal evaluation of
  chatgpt on reasoning, hallucination, and interactivity}.
\newblock \bibinfo{journal}{\emph{arXiv preprint arXiv:2302.04023}}
  (\bibinfo{year}{2023}).
\newblock


\bibitem[Barei{\ss} et~al\mbox{.}(2022)]%
        {codexStudy2022}
\bibfield{author}{\bibinfo{person}{Patrick Barei{\ss}},
  \bibinfo{person}{Beatriz Souza}, \bibinfo{person}{Marcelo d'Amorim}, {and}
  \bibinfo{person}{Michael Pradel}.} \bibinfo{year}{2022}\natexlab{}.
\newblock \showarticletitle{Code Generation Tools (Almost) for Free? {A} Study
  of Few-Shot, Pre-Trained Language Models on Code}.
\newblock \bibinfo{journal}{\emph{CoRR}}  \bibinfo{volume}{abs/2206.01335}
  (\bibinfo{year}{2022}).
\newblock
\urldef\tempurl%
\url{https://doi.org/10.48550/arXiv.2206.01335}
\showDOI{\tempurl}
\showeprint[arXiv]{2206.01335}


\bibitem[B{\"o}hme et~al\mbox{.}(2020)]%
        {bohme2020fuzzing}
\bibfield{author}{\bibinfo{person}{Marcel B{\"o}hme}, \bibinfo{person}{Cristian
  Cadar}, {and} \bibinfo{person}{Abhik Roychoudhury}.}
  \bibinfo{year}{2020}\natexlab{}.
\newblock \showarticletitle{Fuzzing: Challenges and reflections}.
\newblock \bibinfo{journal}{\emph{IEEE Software}} \bibinfo{volume}{38},
  \bibinfo{number}{3} (\bibinfo{year}{2020}), \bibinfo{pages}{79--86}.
\newblock


\bibitem[B{\"o}hme et~al\mbox{.}(2022)]%
        {bohme2022reliability}
\bibfield{author}{\bibinfo{person}{Marcel B{\"o}hme},
  \bibinfo{person}{L{\'a}szl{\'o} Szekeres}, {and} \bibinfo{person}{Jonathan
  Metzman}.} \bibinfo{year}{2022}\natexlab{}.
\newblock \showarticletitle{On the reliability of coverage-based fuzzer
  benchmarking}. In \bibinfo{booktitle}{\emph{Proceedings of the 44th
  International Conference on Software Engineering}}.
  \bibinfo{pages}{1621--1633}.
\newblock


\bibitem[Brown et~al\mbox{.}(2020)]%
        {brown2020gpt3}
\bibfield{author}{\bibinfo{person}{Tom~B. Brown}, \bibinfo{person}{Benjamin
  Mann}, \bibinfo{person}{Nick Ryder}, \bibinfo{person}{Melanie Subbiah},
  \bibinfo{person}{Jared Kaplan}, \bibinfo{person}{Prafulla Dhariwal},
  \bibinfo{person}{Arvind Neelakantan}, \bibinfo{person}{Pranav Shyam},
  \bibinfo{person}{Girish Sastry}, \bibinfo{person}{Amanda Askell},
  \bibinfo{person}{Sandhini Agarwal}, \bibinfo{person}{Ariel Herbert-Voss},
  \bibinfo{person}{Gretchen Krueger}, \bibinfo{person}{Tom Henighan},
  \bibinfo{person}{Rewon Child}, \bibinfo{person}{Aditya Ramesh},
  \bibinfo{person}{Daniel~M. Ziegler}, \bibinfo{person}{Jeffrey Wu},
  \bibinfo{person}{Clemens Winter}, \bibinfo{person}{Christopher Hesse},
  \bibinfo{person}{Mark Chen}, \bibinfo{person}{Eric Sigler},
  \bibinfo{person}{Mateusz Litwin}, \bibinfo{person}{Scott Gray},
  \bibinfo{person}{Benjamin Chess}, \bibinfo{person}{Jack Clark},
  \bibinfo{person}{Christopher Berner}, \bibinfo{person}{Sam McCandlish},
  \bibinfo{person}{Alec Radford}, \bibinfo{person}{Ilya Sutskever}, {and}
  \bibinfo{person}{Dario Amodei}.} \bibinfo{year}{2020}\natexlab{}.
\newblock \bibinfo{title}{Language Models are Few-Shot Learners}.
\newblock
\newblock
\newblock
\shownote{arXiv:2005.14165}.


\bibitem[Bubeck et~al\mbox{.}(2023)]%
        {bubeck2023sparks}
\bibfield{author}{\bibinfo{person}{S{\'e}bastien Bubeck},
  \bibinfo{person}{Varun Chandrasekaran}, \bibinfo{person}{Ronen Eldan},
  \bibinfo{person}{Johannes Gehrke}, \bibinfo{person}{Eric Horvitz},
  \bibinfo{person}{Ece Kamar}, \bibinfo{person}{Peter Lee},
  \bibinfo{person}{Yin~Tat Lee}, \bibinfo{person}{Yuanzhi Li},
  \bibinfo{person}{Scott Lundberg}, {et~al\mbox{.}}}
  \bibinfo{year}{2023}\natexlab{}.
\newblock \showarticletitle{Sparks of artificial general intelligence: Early
  experiments with gpt-4}.
\newblock \bibinfo{journal}{\emph{arXiv preprint arXiv:2303.12712}}
  (\bibinfo{year}{2023}).
\newblock


\bibitem[Bulekov et~al\mbox{.}(2023)]%
        {bulekovno2023grammar}
\bibfield{author}{\bibinfo{person}{Alexander Bulekov}, \bibinfo{person}{Bandan
  Das}, \bibinfo{person}{Stefan Hajnoczi}, {and} \bibinfo{person}{Manuel
  Egele}.} \bibinfo{year}{2023}\natexlab{}.
\newblock \showarticletitle{No Grammar, No Problem: Towards Fuzzing the Linux
  Kernel without System-Call Descriptions}. In
  \bibinfo{booktitle}{\emph{Network and Distributed System Security (NDSS)
  Symposium 2023}}.
\newblock


\bibitem[Chaliasos et~al\mbox{.}(2022)]%
        {chaliasos2022hephaestus}
\bibfield{author}{\bibinfo{person}{Stefanos Chaliasos},
  \bibinfo{person}{Thodoris Sotiropoulos}, \bibinfo{person}{Diomidis
  Spinellis}, \bibinfo{person}{Arthur Gervais}, \bibinfo{person}{Benjamin
  Livshits}, {and} \bibinfo{person}{Dimitris Mitropoulos}.}
  \bibinfo{year}{2022}\natexlab{}.
\newblock \showarticletitle{Finding typing compiler bugs}. In
  \bibinfo{booktitle}{\emph{Proceedings of the 43rd ACM SIGPLAN International
  Conference on Programming Language Design and Implementation}}.
  \bibinfo{pages}{183--198}.
\newblock


\bibitem[Chen et~al\mbox{.}(2020)]%
        {chen2020survey}
\bibfield{author}{\bibinfo{person}{Junjie Chen}, \bibinfo{person}{Jibesh
  Patra}, \bibinfo{person}{Michael Pradel}, \bibinfo{person}{Yingfei Xiong},
  \bibinfo{person}{Hongyu Zhang}, \bibinfo{person}{Dan Hao}, {and}
  \bibinfo{person}{Lu Zhang}.} \bibinfo{year}{2020}\natexlab{}.
\newblock \showarticletitle{A survey of compiler testing}.
\newblock \bibinfo{journal}{\emph{ACM Computing Surveys (CSUR)}}
  \bibinfo{volume}{53}, \bibinfo{number}{1} (\bibinfo{year}{2020}),
  \bibinfo{pages}{1--36}.
\newblock


\bibitem[Chen et~al\mbox{.}(2021a)]%
        {codex}
\bibfield{author}{\bibinfo{person}{Mark Chen}, \bibinfo{person}{Jerry Tworek},
  \bibinfo{person}{Heewoo Jun}, \bibinfo{person}{Qiming Yuan},
  \bibinfo{person}{Henrique Ponde de~Oliveira Pinto}, \bibinfo{person}{Jared
  Kaplan}, \bibinfo{person}{Harri Edwards}, \bibinfo{person}{Yuri Burda},
  \bibinfo{person}{Nicholas Joseph}, \bibinfo{person}{Greg Brockman},
  {et~al\mbox{.}}} \bibinfo{year}{2021}\natexlab{a}.
\newblock \showarticletitle{Evaluating large language models trained on code}.
\newblock \bibinfo{journal}{\emph{arXiv preprint arXiv:2107.03374}}
  (\bibinfo{year}{2021}).
\newblock


\bibitem[Chen et~al\mbox{.}(2021b)]%
        {chen2021polyglot}
\bibfield{author}{\bibinfo{person}{Yongheng Chen}, \bibinfo{person}{Rui Zhong},
  \bibinfo{person}{Hong Hu}, \bibinfo{person}{Hangfan Zhang},
  \bibinfo{person}{Yupeng Yang}, \bibinfo{person}{Dinghao Wu}, {and}
  \bibinfo{person}{Wenke Lee}.} \bibinfo{year}{2021}\natexlab{b}.
\newblock \showarticletitle{One engine to fuzz’em all: Generic language
  processor testing with semantic validation}. In
  \bibinfo{booktitle}{\emph{2021 IEEE Symposium on Security and Privacy (SP)}}.
  IEEE, \bibinfo{pages}{642--658}.
\newblock


\bibitem[Chowdhery et~al\mbox{.}(2022)]%
        {chowdhery2022palm}
\bibfield{author}{\bibinfo{person}{Aakanksha Chowdhery},
  \bibinfo{person}{Sharan Narang}, \bibinfo{person}{Jacob Devlin},
  \bibinfo{person}{Maarten Bosma}, \bibinfo{person}{Gaurav Mishra},
  \bibinfo{person}{Adam Roberts}, \bibinfo{person}{Paul Barham},
  \bibinfo{person}{Hyung~Won Chung}, \bibinfo{person}{Charles Sutton},
  \bibinfo{person}{Sebastian Gehrmann}, \bibinfo{person}{Parker Schuh},
  \bibinfo{person}{Kensen Shi}, \bibinfo{person}{Sasha Tsvyashchenko},
  \bibinfo{person}{Joshua Maynez}, \bibinfo{person}{Abhishek Rao},
  \bibinfo{person}{Parker Barnes}, \bibinfo{person}{Yi Tay},
  \bibinfo{person}{Noam Shazeer}, \bibinfo{person}{Vinodkumar Prabhakaran},
  \bibinfo{person}{Emily Reif}, \bibinfo{person}{Nan Du}, \bibinfo{person}{Ben
  Hutchinson}, \bibinfo{person}{Reiner Pope}, \bibinfo{person}{James Bradbury},
  \bibinfo{person}{Jacob Austin}, \bibinfo{person}{Michael Isard},
  \bibinfo{person}{Guy Gur-Ari}, \bibinfo{person}{Pengcheng Yin},
  \bibinfo{person}{Toju Duke}, \bibinfo{person}{Anselm Levskaya},
  \bibinfo{person}{Sanjay Ghemawat}, \bibinfo{person}{Sunipa Dev},
  \bibinfo{person}{Henryk Michalewski}, \bibinfo{person}{Xavier Garcia},
  \bibinfo{person}{Vedant Misra}, \bibinfo{person}{Kevin Robinson},
  \bibinfo{person}{Liam Fedus}, \bibinfo{person}{Denny Zhou},
  \bibinfo{person}{Daphne Ippolito}, \bibinfo{person}{David Luan},
  \bibinfo{person}{Hyeontaek Lim}, \bibinfo{person}{Barret Zoph},
  \bibinfo{person}{Alexander Spiridonov}, \bibinfo{person}{Ryan Sepassi},
  \bibinfo{person}{David Dohan}, \bibinfo{person}{Shivani Agrawal},
  \bibinfo{person}{Mark Omernick}, \bibinfo{person}{Andrew~M. Dai},
  \bibinfo{person}{Thanumalayan~Sankaranarayana Pillai}, \bibinfo{person}{Marie
  Pellat}, \bibinfo{person}{Aitor Lewkowycz}, \bibinfo{person}{Erica Moreira},
  \bibinfo{person}{Rewon Child}, \bibinfo{person}{Oleksandr Polozov},
  \bibinfo{person}{Katherine Lee}, \bibinfo{person}{Zongwei Zhou},
  \bibinfo{person}{Xuezhi Wang}, \bibinfo{person}{Brennan Saeta},
  \bibinfo{person}{Mark Diaz}, \bibinfo{person}{Orhan Firat},
  \bibinfo{person}{Michele Catasta}, \bibinfo{person}{Jason Wei},
  \bibinfo{person}{Kathy Meier-Hellstern}, \bibinfo{person}{Douglas Eck},
  \bibinfo{person}{Jeff Dean}, \bibinfo{person}{Slav Petrov}, {and}
  \bibinfo{person}{Noah Fiedel}.} \bibinfo{year}{2022}\natexlab{}.
\newblock \bibinfo{title}{PaLM: Scaling Language Modeling with Pathways}.
\newblock
\newblock
\showeprint[arxiv]{2204.02311}~[cs.CL]


\bibitem[Cross et~al\mbox{.}(2017)]%
        {crossOpenQuantumAssembly2017}
\bibfield{author}{\bibinfo{person}{Andrew~W. Cross}, \bibinfo{person}{Lev~S.
  Bishop}, \bibinfo{person}{John~A. Smolin}, {and} \bibinfo{person}{Jay~M.
  Gambetta}.} \bibinfo{year}{2017}\natexlab{}.
\newblock \showarticletitle{Open {{Quantum Assembly Language}}}.
\newblock \bibinfo{journal}{\emph{arXiv:1707.03429 [quant-ph]}}
  (\bibinfo{date}{July} \bibinfo{year}{2017}).
\newblock
\showeprint[arxiv]{1707.03429}~[quant-ph]


\bibitem[Cummins et~al\mbox{.}(2018)]%
        {cummins2018compiler}
\bibfield{author}{\bibinfo{person}{Chris Cummins}, \bibinfo{person}{Pavlos
  Petoumenos}, \bibinfo{person}{Alastair Murray}, {and} \bibinfo{person}{Hugh
  Leather}.} \bibinfo{year}{2018}\natexlab{}.
\newblock \showarticletitle{Compiler fuzzing through deep learning}. In
  \bibinfo{booktitle}{\emph{Proceedings of the 27th ACM SIGSOFT International
  Symposium on Software Testing and Analysis}}. \bibinfo{pages}{95--105}.
\newblock


\bibitem[Deng et~al\mbox{.}(2023)]%
        {deng2023titanfuzz}
\bibfield{author}{\bibinfo{person}{Yinlin Deng},
  \bibinfo{person}{Chunqiu~Steven Xia}, \bibinfo{person}{Haoran Peng},
  \bibinfo{person}{Chenyuan Yang}, {and} \bibinfo{person}{Lingming Zhang}.}
  \bibinfo{year}{2023}\natexlab{}.
\newblock \showarticletitle{Large Language Models are Zero-Shot Fuzzers:
  Fuzzing Deep-Learning Libraries via Large Language Models}. In
  \bibinfo{booktitle}{\emph{Proceedings of the 32nd ACM SIGSOFT International
  Symposium on Software Testing and Analysis}}. \bibinfo{pages}{423--435}.
\newblock


\bibitem[Devlin et~al\mbox{.}(2018)]%
        {devlin2018bert}
\bibfield{author}{\bibinfo{person}{Jacob Devlin}, \bibinfo{person}{Ming-Wei
  Chang}, \bibinfo{person}{Kenton Lee}, {and} \bibinfo{person}{Kristina
  Toutanova}.} \bibinfo{year}{2018}\natexlab{}.
\newblock \showarticletitle{Bert: Pre-training of deep bidirectional
  transformers for language understanding}.
\newblock \bibinfo{journal}{\emph{arXiv preprint arXiv:1810.04805}}
  (\bibinfo{year}{2018}).
\newblock


\bibitem[Even-Mendoza et~al\mbox{.}(2022)]%
        {even2022csmithedge}
\bibfield{author}{\bibinfo{person}{Karine Even-Mendoza},
  \bibinfo{person}{Cristian Cadar}, {and} \bibinfo{person}{Alastair~F
  Donaldson}.} \bibinfo{year}{2022}\natexlab{}.
\newblock \showarticletitle{CsmithEdge: more effective compiler testing by
  handling undefined behaviour less conservatively}.
\newblock \bibinfo{journal}{\emph{Empirical Software Engineering}}
  \bibinfo{volume}{27}, \bibinfo{number}{6} (\bibinfo{year}{2022}),
  \bibinfo{pages}{129}.
\newblock


\bibitem[Even-Mendoza et~al\mbox{.}(2023)]%
        {evenmendoza2023grayc}
\bibfield{author}{\bibinfo{person}{Karine Even-Mendoza},
  \bibinfo{person}{Arindam Sharma}, \bibinfo{person}{Alastair~F. Donaldson},
  {and} \bibinfo{person}{Cristian Cadar}.} \bibinfo{year}{2023}\natexlab{}.
\newblock \showarticletitle{GrayC: Greybox Fuzzing of Compilers and Analysers
  for C} \emph{(\bibinfo{series}{ISSTA 2023})}. \bibinfo{publisher}{Association
  for Computing Machinery}, \bibinfo{address}{New York, NY, USA},
  \bibinfo{pages}{1219–1231}.
\newblock
\showISBNx{9798400702211}
\urldef\tempurl%
\url{https://doi.org/10.1145/3597926.3598130}
\showDOI{\tempurl}


\bibitem[Feng et~al\mbox{.}(2020)]%
        {feng2020codebert}
\bibfield{author}{\bibinfo{person}{Zhangyin Feng}, \bibinfo{person}{Daya Guo},
  \bibinfo{person}{Duyu Tang}, \bibinfo{person}{Nan Duan},
  \bibinfo{person}{Xiaocheng Feng}, \bibinfo{person}{Ming Gong},
  \bibinfo{person}{Linjun Shou}, \bibinfo{person}{Bing Qin},
  \bibinfo{person}{Ting Liu}, \bibinfo{person}{Daxin Jiang}, {and}
  \bibinfo{person}{Ming Zhou}.} \bibinfo{year}{2020}\natexlab{}.
\newblock \bibinfo{title}{CodeBERT: A Pre-Trained Model for Programming and
  Natural Languages}.
\newblock
\newblock
\newblock
\shownote{arXiv:2002.08155}.


\bibitem[Fingerhuth et~al\mbox{.}(2018)]%
        {fingerhuthOpenSourceSoftware2018}
\bibfield{author}{\bibinfo{person}{Mark Fingerhuth},
  \bibinfo{person}{Tom{\'a}{\v s} Babej}, {and} \bibinfo{person}{Peter
  Wittek}.} \bibinfo{year}{2018}\natexlab{}.
\newblock \showarticletitle{Open Source Software in Quantum Computing}.
\newblock \bibinfo{journal}{\emph{PLOS ONE}} \bibinfo{volume}{13},
  \bibinfo{number}{12} (\bibinfo{date}{Dec.} \bibinfo{year}{2018}),
  \bibinfo{pages}{e0208561}.
\newblock
\showISSN{1932-6203}
\urldef\tempurl%
\url{https://doi.org/10.1371/journal.pone.0208561}
\showDOI{\tempurl}


\bibitem[Fried et~al\mbox{.}(2022)]%
        {incoder}
\bibfield{author}{\bibinfo{person}{Daniel Fried}, \bibinfo{person}{Armen
  Aghajanyan}, \bibinfo{person}{Jessy Lin}, \bibinfo{person}{Sida Wang},
  \bibinfo{person}{Eric Wallace}, \bibinfo{person}{Freda Shi},
  \bibinfo{person}{Ruiqi Zhong}, \bibinfo{person}{Wen-tau Yih},
  \bibinfo{person}{Luke Zettlemoyer}, {and} \bibinfo{person}{Mike Lewis}.}
  \bibinfo{year}{2022}\natexlab{}.
\newblock \showarticletitle{Incoder: A generative model for code infilling and
  synthesis}.
\newblock \bibinfo{journal}{\emph{arXiv preprint arXiv:2204.05999}}
  (\bibinfo{year}{2022}).
\newblock


\bibitem[go-fuzz(2023)]%
        {gofuzz}
go-fuzz \bibinfo{year}{2023}\natexlab{}.
\newblock \bibinfo{title}{go-fuzz: randomized testing for Go}.
\newblock
\newblock
\newblock
\shownote{\url{https://github.com/dvyukov/go-fuzz}}.


\bibitem[Godefroid et~al\mbox{.}(2017)]%
        {godefroid2017learn}
\bibfield{author}{\bibinfo{person}{Patrice Godefroid}, \bibinfo{person}{Hila
  Peleg}, {and} \bibinfo{person}{Rishabh Singh}.}
  \bibinfo{year}{2017}\natexlab{}.
\newblock \showarticletitle{Learn\&fuzz: Machine learning for input fuzzing}.
  In \bibinfo{booktitle}{\emph{2017 32nd IEEE/ACM International Conference on
  Automated Software Engineering (ASE)}}. IEEE, \bibinfo{pages}{50--59}.
\newblock


\bibitem[gpt4endpoint(2023)]%
        {gpt4endpoint}
gpt4endpoint \bibinfo{year}{2023}\natexlab{}.
\newblock \bibinfo{title}{Models - GPT-4}.
\newblock
\newblock
\newblock
\shownote{\url{https://platform.openai.com/docs/models/gpt-4}}.


\bibitem[Groce et~al\mbox{.}(2022)]%
        {groce2022making}
\bibfield{author}{\bibinfo{person}{Alex Groce}, \bibinfo{person}{Rijnard van
  Tonder}, \bibinfo{person}{Goutamkumar~Tulajappa Kalburgi}, {and}
  \bibinfo{person}{Claire Le~Goues}.} \bibinfo{year}{2022}\natexlab{}.
\newblock \showarticletitle{Making no-fuss compiler fuzzing effective}. In
  \bibinfo{booktitle}{\emph{Proceedings of the 31st ACM SIGPLAN International
  Conference on Compiler Construction}}. \bibinfo{pages}{194--204}.
\newblock


\bibitem[Gulwani et~al\mbox{.}(2017)]%
        {gulwani2017program}
\bibfield{author}{\bibinfo{person}{Sumit Gulwani}, \bibinfo{person}{Oleksandr
  Polozov}, \bibinfo{person}{Rishabh Singh}, {et~al\mbox{.}}}
  \bibinfo{year}{2017}\natexlab{}.
\newblock \showarticletitle{Program synthesis}.
\newblock \bibinfo{journal}{\emph{Foundations and Trends{\textregistered} in
  Programming Languages}} \bibinfo{volume}{4}, \bibinfo{number}{1-2}
  (\bibinfo{year}{2017}), \bibinfo{pages}{1--119}.
\newblock


\bibitem[Guo et~al\mbox{.}(2022)]%
        {guo2022survey}
\bibfield{author}{\bibinfo{person}{Zhijiang Guo}, \bibinfo{person}{Michael
  Schlichtkrull}, {and} \bibinfo{person}{Andreas Vlachos}.}
  \bibinfo{year}{2022}\natexlab{}.
\newblock \showarticletitle{A survey on automated fact-checking}.
\newblock \bibinfo{journal}{\emph{Transactions of the Association for
  Computational Linguistics}}  \bibinfo{volume}{10} (\bibinfo{year}{2022}),
  \bibinfo{pages}{178--206}.
\newblock


\bibitem[Holler et~al\mbox{.}(2012)]%
        {holler2012fuzzing}
\bibfield{author}{\bibinfo{person}{Christian Holler}, \bibinfo{person}{Kim
  Herzig}, {and} \bibinfo{person}{Andreas Zeller}.}
  \bibinfo{year}{2012}\natexlab{}.
\newblock \showarticletitle{Fuzzing with code fragments}. In
  \bibinfo{booktitle}{\emph{21st USENIX Security Symposium (USENIX Security
  12)}}. \bibinfo{pages}{445--458}.
\newblock


\bibitem[Holtzman et~al\mbox{.}(2019)]%
        {holtzman2019nucleus}
\bibfield{author}{\bibinfo{person}{Ari Holtzman}, \bibinfo{person}{Jan Buys},
  \bibinfo{person}{Li Du}, \bibinfo{person}{Maxwell Forbes}, {and}
  \bibinfo{person}{Yejin Choi}.} \bibinfo{year}{2019}\natexlab{}.
\newblock \bibinfo{title}{The Curious Case of Neural Text Degeneration}.
\newblock
\newblock
\newblock
\shownote{arXiv:1904.09751}.


\bibitem[Jiang et~al\mbox{.}(2020)]%
        {jiang2020cudasmith}
\bibfield{author}{\bibinfo{person}{Bo Jiang}, \bibinfo{person}{Xiaoyan Wang},
  \bibinfo{person}{Wing~Kwong Chan}, \bibinfo{person}{TH Tse},
  \bibinfo{person}{Na Li}, \bibinfo{person}{Yongfeng Yin}, {and}
  \bibinfo{person}{Zhenyu Zhang}.} \bibinfo{year}{2020}\natexlab{}.
\newblock \showarticletitle{Cudasmith: A fuzzer for CUDA compilers}. In
  \bibinfo{booktitle}{\emph{2020 IEEE 44th Annual Computers, Software, and
  Applications Conference (COMPSAC)}}. IEEE, \bibinfo{pages}{861--871}.
\newblock


\bibitem[jsfunfuzz(2017)]%
        {jsfunfuzz}
jsfunfuzz \bibinfo{year}{2017}\natexlab{}.
\newblock \bibinfo{title}{Introducing jsfunfuzz}.
\newblock
\newblock
\newblock
\shownote{\url{https://www.squarefree.com/2007/08/02/introducing-jsfunfuzz/}}.


\bibitem[Kaplan et~al\mbox{.}(2020)]%
        {kaplan2020scaling}
\bibfield{author}{\bibinfo{person}{Jared Kaplan}, \bibinfo{person}{Sam
  McCandlish}, \bibinfo{person}{Tom Henighan}, \bibinfo{person}{Tom~B Brown},
  \bibinfo{person}{Benjamin Chess}, \bibinfo{person}{Rewon Child},
  \bibinfo{person}{Scott Gray}, \bibinfo{person}{Alec Radford},
  \bibinfo{person}{Jeffrey Wu}, {and} \bibinfo{person}{Dario Amodei}.}
  \bibinfo{year}{2020}\natexlab{}.
\newblock \showarticletitle{Scaling laws for neural language models}.
\newblock \bibinfo{journal}{\emph{arXiv preprint arXiv:2001.08361}}
  (\bibinfo{year}{2020}).
\newblock


\bibitem[Klees et~al\mbox{.}(2018)]%
        {kleesEvaluatingFuzzTesting2018}
\bibfield{author}{\bibinfo{person}{George Klees}, \bibinfo{person}{Andrew
  Ruef}, \bibinfo{person}{Benji Cooper}, \bibinfo{person}{Shiyi Wei}, {and}
  \bibinfo{person}{Michael Hicks}.} \bibinfo{year}{2018}\natexlab{}.
\newblock \showarticletitle{Evaluating {{Fuzz Testing}}}. In
  \bibinfo{booktitle}{\emph{Proceedings of the 2018 {{ACM SIGSAC Conference}}
  on {{Computer}} and {{Communications Security}}}}
  \emph{(\bibinfo{series}{{{CCS}} '18})}. \bibinfo{publisher}{{Association for
  Computing Machinery}}, \bibinfo{address}{{New York, NY, USA}},
  \bibinfo{pages}{2123--2138}.
\newblock
\showISBNx{978-1-4503-5693-0}
\urldef\tempurl%
\url{https://doi.org/10.1145/3243734.3243804}
\showDOI{\tempurl}


\bibitem[Lee et~al\mbox{.}(2020)]%
        {lee2020montage}
\bibfield{author}{\bibinfo{person}{Suyoung Lee}, \bibinfo{person}{HyungSeok
  Han}, \bibinfo{person}{Sang~Kil Cha}, {and} \bibinfo{person}{Sooel Son}.}
  \bibinfo{year}{2020}\natexlab{}.
\newblock \showarticletitle{Montage: A Neural Network Language
  $\{$Model-Guided$\}$$\{$JavaScript$\}$ Engine Fuzzer}. In
  \bibinfo{booktitle}{\emph{29th USENIX Security Symposium (USENIX Security
  20)}}. \bibinfo{pages}{2613--2630}.
\newblock


\bibitem[Lehmann and Pradel(2018)]%
        {fse2018}
\bibfield{author}{\bibinfo{person}{Daniel Lehmann} {and}
  \bibinfo{person}{Michael Pradel}.} \bibinfo{year}{2018}\natexlab{}.
\newblock \showarticletitle{Feedback-directed differential testing of
  interactive debuggers}. In \bibinfo{booktitle}{\emph{{ESEC/SIGSOFT} {FSE}}}.
  \bibinfo{pages}{610--620}.
\newblock


\bibitem[Lemieux et~al\mbox{.}(2023)]%
        {lemieux2023codamosa}
\bibfield{author}{\bibinfo{person}{Caroline Lemieux},
  \bibinfo{person}{Jeevana~Priya Inala}, \bibinfo{person}{Shuvendu~K Lahiri},
  {and} \bibinfo{person}{Siddhartha Sen}.} \bibinfo{year}{2023}\natexlab{}.
\newblock \showarticletitle{CODAMOSA: Escaping Coverage Plateaus in Test
  Generation with Pre-trained Large Language Models}. In
  \bibinfo{booktitle}{\emph{45th International Conference on Software
  Engineering}}.
\newblock


\bibitem[Lewis et~al\mbox{.}(2019)]%
        {lewis2019bart}
\bibfield{author}{\bibinfo{person}{Mike Lewis}, \bibinfo{person}{Yinhan Liu},
  \bibinfo{person}{Naman Goyal}, \bibinfo{person}{Marjan Ghazvininejad},
  \bibinfo{person}{Abdelrahman Mohamed}, \bibinfo{person}{Omer Levy},
  \bibinfo{person}{Ves Stoyanov}, {and} \bibinfo{person}{Luke Zettlemoyer}.}
  \bibinfo{year}{2019}\natexlab{}.
\newblock \showarticletitle{Bart: Denoising sequence-to-sequence pre-training
  for natural language generation, translation, and comprehension}.
\newblock \bibinfo{journal}{\emph{arXiv preprint arXiv:1910.13461}}
  (\bibinfo{year}{2019}).
\newblock


\bibitem[Li et~al\mbox{.}(2023)]%
        {li2023starcoder}
\bibfield{author}{\bibinfo{person}{Raymond Li}, \bibinfo{person}{Loubna~Ben
  Allal}, \bibinfo{person}{Yangtian Zi}, \bibinfo{person}{Niklas Muennighoff},
  \bibinfo{person}{Denis Kocetkov}, \bibinfo{person}{Chenghao Mou},
  \bibinfo{person}{Marc Marone}, \bibinfo{person}{Christopher Akiki},
  \bibinfo{person}{Jia Li}, \bibinfo{person}{Jenny Chim}, {et~al\mbox{.}}}
  \bibinfo{year}{2023}\natexlab{}.
\newblock \showarticletitle{StarCoder: may the source be with you!}
\newblock \bibinfo{journal}{\emph{arXiv preprint arXiv:2305.06161}}
  (\bibinfo{year}{2023}).
\newblock


\bibitem[Li and Liang(2021)]%
        {li2021prefix}
\bibfield{author}{\bibinfo{person}{Xiang~Lisa Li} {and} \bibinfo{person}{Percy
  Liang}.} \bibinfo{year}{2021}\natexlab{}.
\newblock \showarticletitle{Prefix-tuning: Optimizing continuous prompts for
  generation}.
\newblock \bibinfo{journal}{\emph{arXiv preprint arXiv:2101.00190}}
  (\bibinfo{year}{2021}).
\newblock


\bibitem[libFuzzer(2023)]%
        {libfuzzer}
libFuzzer \bibinfo{year}{2023}\natexlab{}.
\newblock \bibinfo{title}{libFuzzer – a library for coverage-guided fuzz
  testing.}
\newblock
\newblock
\newblock
\shownote{\url{https://llvm.org/docs/LibFuzzer.html}}.


\bibitem[Lidbury et~al\mbox{.}(2015)]%
        {lidbury2015clsmith}
\bibfield{author}{\bibinfo{person}{Christopher Lidbury},
  \bibinfo{person}{Andrei Lascu}, \bibinfo{person}{Nathan Chong}, {and}
  \bibinfo{person}{Alastair~F Donaldson}.} \bibinfo{year}{2015}\natexlab{}.
\newblock \showarticletitle{Many-core compiler fuzzing}.
\newblock \bibinfo{journal}{\emph{ACM SIGPLAN Notices}} \bibinfo{volume}{50},
  \bibinfo{number}{6} (\bibinfo{year}{2015}), \bibinfo{pages}{65--76}.
\newblock


\bibitem[Liu et~al\mbox{.}(2023)]%
        {liu2023nnsmith}
\bibfield{author}{\bibinfo{person}{Jiawei Liu}, \bibinfo{person}{Jinkun Lin},
  \bibinfo{person}{Fabian Ruffy}, \bibinfo{person}{Cheng Tan},
  \bibinfo{person}{Jinyang Li}, \bibinfo{person}{Aurojit Panda}, {and}
  \bibinfo{person}{Lingming Zhang}.} \bibinfo{year}{2023}\natexlab{}.
\newblock \showarticletitle{Nnsmith: Generating diverse and valid test cases
  for deep learning compilers}. In \bibinfo{booktitle}{\emph{Proceedings of the
  28th ACM International Conference on Architectural Support for Programming
  Languages and Operating Systems, Volume 2}}. \bibinfo{pages}{530--543}.
\newblock


\bibitem[Liu et~al\mbox{.}(2022)]%
        {liu2022coverage}
\bibfield{author}{\bibinfo{person}{Jiawei Liu}, \bibinfo{person}{Yuxiang Wei},
  \bibinfo{person}{Sen Yang}, \bibinfo{person}{Yinlin Deng}, {and}
  \bibinfo{person}{Lingming Zhang}.} \bibinfo{year}{2022}\natexlab{}.
\newblock \showarticletitle{Coverage-guided tensor compiler fuzzing with joint
  ir-pass mutation}.
\newblock \bibinfo{journal}{\emph{Proceedings of the ACM on Programming
  Languages}} \bibinfo{volume}{6}, \bibinfo{number}{OOPSLA1}
  (\bibinfo{year}{2022}), \bibinfo{pages}{1--26}.
\newblock


\bibitem[Liu et~al\mbox{.}(2021)]%
        {Liu2021a}
\bibfield{author}{\bibinfo{person}{Pengfei Liu}, \bibinfo{person}{Weizhe Yuan},
  \bibinfo{person}{Jinlan Fu}, \bibinfo{person}{Zhengbao Jiang},
  \bibinfo{person}{Hiroaki Hayashi}, {and} \bibinfo{person}{Graham Neubig}.}
  \bibinfo{year}{2021}\natexlab{}.
\newblock \showarticletitle{Pre-train, Prompt, and Predict: {A} Systematic
  Survey of Prompting Methods in Natural Language Processing}.
\newblock \bibinfo{journal}{\emph{CoRR}}  \bibinfo{volume}{abs/2107.13586}
  (\bibinfo{year}{2021}).
\newblock
\showeprint[arXiv]{2107.13586}
\urldef\tempurl%
\url{https://arxiv.org/abs/2107.13586}
\showURL{%
\tempurl}


\bibitem[Liu et~al\mbox{.}(2019)]%
        {liu2019deepfuzz}
\bibfield{author}{\bibinfo{person}{Xiao Liu}, \bibinfo{person}{Xiaoting Li},
  \bibinfo{person}{Rupesh Prajapati}, {and} \bibinfo{person}{Dinghao Wu}.}
  \bibinfo{year}{2019}\natexlab{}.
\newblock \showarticletitle{Deepfuzz: Automatic generation of syntax valid c
  programs for fuzz testing}. In \bibinfo{booktitle}{\emph{Proceedings of the
  AAAI Conference on Artificial Intelligence}}, Vol.~\bibinfo{volume}{33}.
  \bibinfo{pages}{1044--1051}.
\newblock


\bibitem[Livinskii et~al\mbox{.}(2020)]%
        {livinskii2020yarpgen}
\bibfield{author}{\bibinfo{person}{Vsevolod Livinskii}, \bibinfo{person}{Dmitry
  Babokin}, {and} \bibinfo{person}{John Regehr}.}
  \bibinfo{year}{2020}\natexlab{}.
\newblock \showarticletitle{Random testing for C and C++ compilers with
  YARPGen}.
\newblock \bibinfo{journal}{\emph{Proceedings of the ACM on Programming
  Languages}} \bibinfo{volume}{4}, \bibinfo{number}{OOPSLA}
  (\bibinfo{year}{2020}), \bibinfo{pages}{1--25}.
\newblock


\bibitem[M. Zalewski(2016)]%
        {afl}
M. Zalewski \bibinfo{year}{2016}\natexlab{}.
\newblock \bibinfo{title}{American Fuzzy Lop - Whitepaper}.
\newblock
\newblock
\newblock
\shownote{\url{https://lcamtuf.coredump.cx/afl/technical_details.txt}}.


\bibitem[Ma(2023)]%
        {ma2023survey}
\bibfield{author}{\bibinfo{person}{Haoyang Ma}.}
  \bibinfo{year}{2023}\natexlab{}.
\newblock \showarticletitle{A Survey of Modern Compiler Fuzzing}.
\newblock \bibinfo{journal}{\emph{arXiv preprint arXiv:2306.06884}}
  (\bibinfo{year}{2023}).
\newblock


\bibitem[Mann and Whitney(1947)]%
        {mann1947test}
\bibfield{author}{\bibinfo{person}{Henry~B Mann} {and}
  \bibinfo{person}{Donald~R Whitney}.} \bibinfo{year}{1947}\natexlab{}.
\newblock \showarticletitle{On a test of whether one of two random variables is
  stochastically larger than the other}.
\newblock \bibinfo{journal}{\emph{The annals of mathematical statistics}}
  (\bibinfo{year}{1947}), \bibinfo{pages}{50--60}.
\newblock


\bibitem[Mansur et~al\mbox{.}(2021)]%
        {DBLP:conf/sigsoft/MansurCW21}
\bibfield{author}{\bibinfo{person}{Muhammad~Numair Mansur},
  \bibinfo{person}{Maria Christakis}, {and} \bibinfo{person}{Valentin
  W{\"{u}}stholz}.} \bibinfo{year}{2021}\natexlab{}.
\newblock \showarticletitle{Metamorphic testing of Datalog engines}. In
  \bibinfo{booktitle}{\emph{{ESEC/FSE} '21: 29th {ACM} Joint European Software
  Engineering Conference and Symposium on the Foundations of Software
  Engineering}}. \bibinfo{pages}{639--650}.
\newblock
\urldef\tempurl%
\url{https://doi.org/10.1145/3468264.3468573}
\showDOI{\tempurl}


\bibitem[Nie et~al\mbox{.}(2023)]%
        {nie2023teco}
\bibfield{author}{\bibinfo{person}{Pengyu Nie}, \bibinfo{person}{Rahul
  Banerjee}, \bibinfo{person}{Junyi~Jessy Li}, \bibinfo{person}{Raymond~J.
  Mooney}, {and} \bibinfo{person}{Milos Gligoric}.}
  \bibinfo{year}{2023}\natexlab{}.
\newblock \showarticletitle{Learning Deep Semantics for Test Completion}. In
  \bibinfo{booktitle}{\emph{45th International Conference on Software
  Engineering}}.
\newblock


\bibitem[OpenAI(2023)]%
        {openai2023gpt4}
\bibfield{author}{\bibinfo{person}{OpenAI}.} \bibinfo{year}{2023}\natexlab{}.
\newblock \bibinfo{title}{GPT-4 Technical Report}.
\newblock
\newblock
\showeprint[arxiv]{2303.08774}~[cs.CL]


\bibitem[Ouyang et~al\mbox{.}(2022)]%
        {ouyang2022instructgpt}
\bibfield{author}{\bibinfo{person}{Long Ouyang}, \bibinfo{person}{Jeffrey Wu},
  \bibinfo{person}{Xu Jiang}, \bibinfo{person}{Diogo Almeida},
  \bibinfo{person}{Carroll Wainwright}, \bibinfo{person}{Pamela Mishkin},
  \bibinfo{person}{Chong Zhang}, \bibinfo{person}{Sandhini Agarwal},
  \bibinfo{person}{Katarina Slama}, \bibinfo{person}{Alex Ray},
  {et~al\mbox{.}}} \bibinfo{year}{2022}\natexlab{}.
\newblock \showarticletitle{Training language models to follow instructions
  with human feedback}.
\newblock \bibinfo{journal}{\emph{Advances in Neural Information Processing
  Systems}}  \bibinfo{volume}{35} (\bibinfo{year}{2022}),
  \bibinfo{pages}{27730--27744}.
\newblock


\bibitem[Paltenghi and Pradel(2022)]%
        {oopsla2022}
\bibfield{author}{\bibinfo{person}{Matteo Paltenghi} {and}
  \bibinfo{person}{Michael Pradel}.} \bibinfo{year}{2022}\natexlab{}.
\newblock \showarticletitle{Bugs in Quantum computing platforms: an empirical
  study}.
\newblock \bibinfo{journal}{\emph{Proc. {ACM} Program. Lang.}}
  \bibinfo{volume}{6}, \bibinfo{number}{{OOPSLA}} (\bibinfo{year}{2022}),
  \bibinfo{pages}{1--27}.
\newblock
\urldef\tempurl%
\url{https://doi.org/10.1145/3527330}
\showDOI{\tempurl}


\bibitem[Paltenghi and Pradel(2023)]%
        {paltenghiMorphQMetamorphicTesting2023}
\bibfield{author}{\bibinfo{person}{Matteo Paltenghi} {and}
  \bibinfo{person}{Michael Pradel}.} \bibinfo{year}{2023}\natexlab{}.
\newblock \showarticletitle{{{MorphQ}}: {{Metamorphic Testing}} of the {{Qiskit
  Quantum Computing Platform}}}. In \bibinfo{booktitle}{\emph{2023
  {{IEEE}}/{{ACM}} 45th {{International Conference}} on {{Software
  Engineering}} ({{ICSE}})}}. \bibinfo{publisher}{{IEEE Computer Society}},
  \bibinfo{pages}{2413--2424}.
\newblock
\showISBNx{978-1-66545-701-9}
\urldef\tempurl%
\url{https://doi.org/10.1109/ICSE48619.2023.00202}
\showDOI{\tempurl}


\bibitem[Park et~al\mbox{.}(2021)]%
        {park2021generative}
\bibfield{author}{\bibinfo{person}{Jiwon Park}, \bibinfo{person}{Dominik
  Winterer}, \bibinfo{person}{Chengyu Zhang}, {and} \bibinfo{person}{Zhendong
  Su}.} \bibinfo{year}{2021}\natexlab{}.
\newblock \showarticletitle{Generative type-aware mutation for testing SMT
  solvers}.
\newblock \bibinfo{journal}{\emph{Proceedings of the ACM on Programming
  Languages}} \bibinfo{volume}{5}, \bibinfo{number}{OOPSLA}
  (\bibinfo{year}{2021}), \bibinfo{pages}{1--19}.
\newblock


\bibitem[Patra and Pradel(2016)]%
        {patra2016learning}
\bibfield{author}{\bibinfo{person}{Jibesh Patra} {and} \bibinfo{person}{Michael
  Pradel}.} \bibinfo{year}{2016}\natexlab{}.
\newblock \showarticletitle{Learning to fuzz: Application-independent fuzz
  testing with probabilistic, generative models of input data}.
\newblock  (\bibinfo{year}{2016}).
\newblock


\bibitem[PyTorch(2023)]%
        {PyTorch}
PyTorch \bibinfo{year}{2023}\natexlab{}.
\newblock \bibinfo{title}{PyTorch}.
\newblock
\newblock
\newblock
\shownote{\url{http://pytorch.org}}.


\bibitem[Qin and Eisner(2021)]%
        {qin2021learning}
\bibfield{author}{\bibinfo{person}{Guanghui Qin} {and} \bibinfo{person}{Jason
  Eisner}.} \bibinfo{year}{2021}\natexlab{}.
\newblock \showarticletitle{Learning How to Ask: Querying LMs with Mixtures of
  Soft Prompts}. In \bibinfo{booktitle}{\emph{Proceedings of the 2021
  Conference of the North American Chapter of the Association for Computational
  Linguistics: Human Language Technologies (NAACL-HLT)}}.
\newblock


\bibitem[Radford et~al\mbox{.}(2018)]%
        {radford2018improving}
\bibfield{author}{\bibinfo{person}{Alec Radford}, \bibinfo{person}{Karthik
  Narasimhan}, \bibinfo{person}{Tim Salimans}, \bibinfo{person}{Ilya
  Sutskever}, {et~al\mbox{.}}} \bibinfo{year}{2018}\natexlab{}.
\newblock \showarticletitle{Improving language understanding by generative
  pre-training}.
\newblock  (\bibinfo{year}{2018}).
\newblock


\bibitem[Schick and Sch{\"u}tze(2020)]%
        {schick2020exploiting}
\bibfield{author}{\bibinfo{person}{Timo Schick} {and} \bibinfo{person}{Hinrich
  Sch{\"u}tze}.} \bibinfo{year}{2020}\natexlab{}.
\newblock \showarticletitle{Exploiting cloze questions for few shot text
  classification and natural language inference}.
\newblock \bibinfo{journal}{\emph{arXiv preprint arXiv:2001.07676}}
  (\bibinfo{year}{2020}).
\newblock


\bibitem[Schulman et~al\mbox{.}(2022)]%
        {chatgpt}
\bibfield{author}{\bibinfo{person}{John Schulman}, \bibinfo{person}{Barret
  Zoph}, \bibinfo{person}{Jacob~Hilton Christina~Kim}, \bibinfo{person}{Jacob
  Menick}, \bibinfo{person}{Jiayi Weng}, \bibinfo{person}{Juan Felipe~Ceron
  Uribe}, \bibinfo{person}{Liam Fedus}, \bibinfo{person}{Luke Metz},
  \bibinfo{person}{Michael Pokorny}, \bibinfo{person}{Rapha~Gontijo Lopes},
  \bibinfo{person}{Shengjia Zhao}, \bibinfo{person}{Arun Vijayvergiya},
  \bibinfo{person}{Eric Sigler}, \bibinfo{person}{Adam Perelman},
  \bibinfo{person}{Chelsea Voss}, \bibinfo{person}{Mike Heaton},
  \bibinfo{person}{Joel Parish}, \bibinfo{person}{Dave Cummings},
  \bibinfo{person}{Rajeev Nayak}, \bibinfo{person}{Valerie Balcom},
  \bibinfo{person}{David Schnurr}, \bibinfo{person}{Tomer Kaftan},
  \bibinfo{person}{Chris Hallacy}, \bibinfo{person}{Nicholas Turley},
  \bibinfo{person}{Noah Deutsch}, \bibinfo{person}{Vik Goel},
  \bibinfo{person}{Jonathan Ward}, \bibinfo{person}{Aris Konstantinidis},
  \bibinfo{person}{Wojciech Zaremba}, \bibinfo{person}{Long Ouyang},
  \bibinfo{person}{Leonard Bogdonoff}, \bibinfo{person}{Joshua Gross},
  \bibinfo{person}{David Medina}, \bibinfo{person}{Sarah Yoo},
  \bibinfo{person}{Teddy Lee}, \bibinfo{person}{Ryan Lowe},
  \bibinfo{person}{Dan Mossing}, \bibinfo{person}{Joost Huizinga},
  \bibinfo{person}{Roger Jiang}, \bibinfo{person}{Carroll Wainwright},
  \bibinfo{person}{Diogo Almeida}, \bibinfo{person}{Steph Lin},
  \bibinfo{person}{Marvin Zhang}, \bibinfo{person}{Kai Xiao},
  \bibinfo{person}{Katarina Slama}, \bibinfo{person}{Steven Bills},
  \bibinfo{person}{Alex Gray}, \bibinfo{person}{Jan Leike},
  \bibinfo{person}{Jakub Pachocki}, \bibinfo{person}{Phil Tillet},
  \bibinfo{person}{Shantanu Jain}, \bibinfo{person}{Greg Brockman}, {and}
  \bibinfo{person}{Nick Ryder}.} \bibinfo{year}{2022}\natexlab{}.
\newblock \showarticletitle{ChatGPT: Optimizing Language Models for Dialogue}.
\newblock  (\bibinfo{year}{2022}).
\newblock
\newblock
\shownote{\url{https://openai.com/blog/chatgpt/}}.


\bibitem[Schäfer et~al\mbox{.}(2023)]%
        {schafer2023testpilot}
\bibfield{author}{\bibinfo{person}{Max Schäfer}, \bibinfo{person}{Sarah Nadi},
  \bibinfo{person}{Aryaz Eghbali}, {and} \bibinfo{person}{Frank Tip}.}
  \bibinfo{year}{2023}\natexlab{}.
\newblock \bibinfo{title}{Adaptive Test Generation Using a Large Language
  Model}.
\newblock
\newblock
\showeprint[arxiv]{2302.06527}~[cs.SE]


\bibitem[Shi et~al\mbox{.}(2022)]%
        {shi2022tf}
\bibfield{author}{\bibinfo{person}{Kensen Shi}, \bibinfo{person}{David Bieber},
  {and} \bibinfo{person}{Rishabh Singh}.} \bibinfo{year}{2022}\natexlab{}.
\newblock \showarticletitle{Tf-coder: Program synthesis for tensor
  manipulations}.
\newblock \bibinfo{journal}{\emph{ACM Transactions on Programming Languages and
  Systems (TOPLAS)}} \bibinfo{volume}{44}, \bibinfo{number}{2}
  (\bibinfo{year}{2022}), \bibinfo{pages}{1--36}.
\newblock


\bibitem[Shin et~al\mbox{.}(2020)]%
        {shin2020autoprompt}
\bibfield{author}{\bibinfo{person}{Taylor Shin}, \bibinfo{person}{Yasaman
  Razeghi}, \bibinfo{person}{Robert~L Logan~IV}, \bibinfo{person}{Eric
  Wallace}, {and} \bibinfo{person}{Sameer Singh}.}
  \bibinfo{year}{2020}\natexlab{}.
\newblock \showarticletitle{Autoprompt: Eliciting knowledge from language
  models with automatically generated prompts}.
\newblock \bibinfo{journal}{\emph{arXiv preprint arXiv:2010.15980}}
  (\bibinfo{year}{2020}).
\newblock


\bibitem[Sutton et~al\mbox{.}(2007)]%
        {SuttonFuzzingBook}
\bibfield{author}{\bibinfo{person}{Michael Sutton}, \bibinfo{person}{Adam
  Greene}, {and} \bibinfo{person}{Pedram Amini}.}
  \bibinfo{year}{2007}\natexlab{}.
\newblock \bibinfo{booktitle}{\emph{Fuzzing: Brute Force Vulnerability
  Discovery}}.
\newblock \bibinfo{publisher}{Addison-Wesley Professional}.
\newblock


\bibitem[syzkaller(2023)]%
        {syzkaller}
syzkaller \bibinfo{year}{2023}\natexlab{}.
\newblock \bibinfo{title}{syzkaller - kernel fuzzer}.
\newblock
\newblock
\newblock
\shownote{\url{https://github.com/google/syzkaller}}.


\bibitem[Tam et~al\mbox{.}(2021)]%
        {tam2021improving}
\bibfield{author}{\bibinfo{person}{Derek Tam}, \bibinfo{person}{Rakesh~R
  Menon}, \bibinfo{person}{Mohit Bansal}, \bibinfo{person}{Shashank
  Srivastava}, {and} \bibinfo{person}{Colin Raffel}.}
  \bibinfo{year}{2021}\natexlab{}.
\newblock \showarticletitle{Improving and simplifying pattern exploiting
  training}.
\newblock \bibinfo{journal}{\emph{arXiv preprint arXiv:2103.11955}}
  (\bibinfo{year}{2021}).
\newblock


\bibitem[TensorFlow(2023)]%
        {Tensorflow}
TensorFlow \bibinfo{year}{2023}\natexlab{}.
\newblock \bibinfo{title}{TensorFlow}.
\newblock
\newblock
\newblock
\shownote{\url{https://www.tensorflow.org}}.


\bibitem[Vaswani et~al\mbox{.}(2017)]%
        {vaswani2017attention}
\bibfield{author}{\bibinfo{person}{Ashish Vaswani}, \bibinfo{person}{Noam
  Shazeer}, \bibinfo{person}{Niki Parmar}, \bibinfo{person}{Jakob Uszkoreit},
  \bibinfo{person}{Llion Jones}, \bibinfo{person}{Aidan~N Gomez},
  \bibinfo{person}{{\L}ukasz Kaiser}, {and} \bibinfo{person}{Illia
  Polosukhin}.} \bibinfo{year}{2017}\natexlab{}.
\newblock \showarticletitle{Attention is all you need}.
\newblock \bibinfo{journal}{\emph{Advances in neural information processing
  systems}}  \bibinfo{volume}{30} (\bibinfo{year}{2017}).
\newblock


\bibitem[Vikram et~al\mbox{.}(2023)]%
        {vikram2023propertytest}
\bibfield{author}{\bibinfo{person}{Vasudev Vikram}, \bibinfo{person}{Caroline
  Lemieux}, {and} \bibinfo{person}{Rohan Padhye}.}
  \bibinfo{year}{2023}\natexlab{}.
\newblock \showarticletitle{Can Large Language Models Write Good Property-Based
  Tests?}
\newblock \bibinfo{journal}{\emph{arXiv preprint arXiv:2307.04346}}
  (\bibinfo{year}{2023}).
\newblock


\bibitem[Wang et~al\mbox{.}(2022)]%
        {wang2022noprompt}
\bibfield{author}{\bibinfo{person}{Chaozheng Wang}, \bibinfo{person}{Yuanhang
  Yang}, \bibinfo{person}{Cuiyun Gao}, \bibinfo{person}{Yun Peng},
  \bibinfo{person}{Hongyu Zhang}, {and} \bibinfo{person}{Michael~R Lyu}.}
  \bibinfo{year}{2022}\natexlab{}.
\newblock \showarticletitle{No more fine-tuning? an experimental evaluation of
  prompt tuning in code intelligence}. In \bibinfo{booktitle}{\emph{Proceedings
  of the 30th ACM Joint European Software Engineering Conference and Symposium
  on the Foundations of Software Engineering}}. \bibinfo{pages}{382--394}.
\newblock


\bibitem[Wei et~al\mbox{.}(2022)]%
        {wei2022free}
\bibfield{author}{\bibinfo{person}{Anjiang Wei}, \bibinfo{person}{Yinlin Deng},
  \bibinfo{person}{Chenyuan Yang}, {and} \bibinfo{person}{Lingming Zhang}.}
  \bibinfo{year}{2022}\natexlab{}.
\newblock \showarticletitle{Free lunch for testing: Fuzzing deep-learning
  libraries from open source}. In \bibinfo{booktitle}{\emph{Proceedings of the
  44th International Conference on Software Engineering}}.
  \bibinfo{pages}{995--1007}.
\newblock


\bibitem[Winterer et~al\mbox{.}(2020a)]%
        {winterer-zhang-su-oopsla2020}
\bibfield{author}{\bibinfo{person}{Dominik Winterer}, \bibinfo{person}{Chengyu
  Zhang}, {and} \bibinfo{person}{Zhendong Su}.}
  \bibinfo{year}{2020}\natexlab{a}.
\newblock \showarticletitle{On the unusual effectiveness of type-aware operator
  mutations for testing {SMT} solvers}.
\newblock \bibinfo{journal}{\emph{Proc. {ACM} Program. Lang.}}
  \bibinfo{volume}{4}, \bibinfo{number}{{OOPSLA}} (\bibinfo{year}{2020}),
  \bibinfo{pages}{193:1--193:25}.
\newblock


\bibitem[Winterer et~al\mbox{.}(2020b)]%
        {winterer-zhang-su-pldi2020}
\bibfield{author}{\bibinfo{person}{Dominik Winterer}, \bibinfo{person}{Chengyu
  Zhang}, {and} \bibinfo{person}{Zhendong Su}.}
  \bibinfo{year}{2020}\natexlab{b}.
\newblock \showarticletitle{Validating SMT Solvers via Semantic Fusion}. In
  \bibinfo{booktitle}{\emph{Proceedings of the 41st ACM SIGPLAN Conference on
  Programming Language Design and Implementation}}. \bibinfo{pages}{718–730}.
\newblock


\bibitem[Xia and Zhang(2023)]%
        {xia2023chatrepair}
\bibfield{author}{\bibinfo{person}{Chunqiu~Steven Xia} {and}
  \bibinfo{person}{Lingming Zhang}.} \bibinfo{year}{2023}\natexlab{}.
\newblock \showarticletitle{Keep the Conversation Going: Fixing 162 out of 337
  bugs for \$0.42 each using ChatGPT}.
\newblock \bibinfo{journal}{\emph{arXiv preprint arXiv:2304.00385}}
  (\bibinfo{year}{2023}).
\newblock


\bibitem[Xu et~al\mbox{.}(2022)]%
        {xu2022systematic}
\bibfield{author}{\bibinfo{person}{Frank~F. Xu}, \bibinfo{person}{Uri Alon},
  \bibinfo{person}{Graham Neubig}, {and} \bibinfo{person}{Vincent~Josua
  Hellendoorn}.} \bibinfo{year}{2022}\natexlab{}.
\newblock \showarticletitle{A Systematic Evaluation of Large Language Models of
  Code}. In \bibinfo{booktitle}{\emph{Proceedings of the 6th ACM SIGPLAN
  International Symposium on Machine Programming}} (San Diego, CA, USA)
  \emph{(\bibinfo{series}{MAPS 2022})}. \bibinfo{publisher}{Association for
  Computing Machinery}, \bibinfo{address}{New York, NY, USA},
  \bibinfo{pages}{1–10}.
\newblock


\bibitem[Yang et~al\mbox{.}(2011)]%
        {yang2011csmith}
\bibfield{author}{\bibinfo{person}{Xuejun Yang}, \bibinfo{person}{Yang Chen},
  \bibinfo{person}{Eric Eide}, {and} \bibinfo{person}{John Regehr}.}
  \bibinfo{year}{2011}\natexlab{}.
\newblock \showarticletitle{Finding and understanding bugs in C compilers}. In
  \bibinfo{booktitle}{\emph{Proceedings of the 32nd ACM SIGPLAN conference on
  Programming language design and implementation}}. \bibinfo{pages}{283--294}.
\newblock


\bibitem[Yuan et~al\mbox{.}(2023)]%
        {yuan2023manual}
\bibfield{author}{\bibinfo{person}{Zhiqiang Yuan}, \bibinfo{person}{Yiling
  Lou}, \bibinfo{person}{Mingwei Liu}, \bibinfo{person}{Shiji Ding},
  \bibinfo{person}{Kaixin Wang}, \bibinfo{person}{Yixuan Chen}, {and}
  \bibinfo{person}{Xin Peng}.} \bibinfo{year}{2023}\natexlab{}.
\newblock \bibinfo{title}{No More Manual Tests? Evaluating and Improving
  ChatGPT for Unit Test Generation}.
\newblock
\newblock
\showeprint[arxiv]{2305.04207}~[cs.SE]


\bibitem[Yue~Wang and Hoi(2021)]%
        {wang2021codet5}
\bibfield{author}{\bibinfo{person}{Shafiq~Joty Yue~Wang, Weishi~Wang} {and}
  \bibinfo{person}{Steven~C.H. Hoi}.} \bibinfo{year}{2021}\natexlab{}.
\newblock \showarticletitle{CodeT5: Identifier-aware Unified Pre-trained
  Encoder-Decoder Models for Code Understanding and Generation}. In
  \bibinfo{booktitle}{\emph{Proceedings of the 2021 Conference on Empirical
  Methods in Natural Language Processing, EMNLP 2021}}.
\newblock


\bibitem[Zeller et~al\mbox{.}(2019)]%
        {zeller2019fuzzing}
\bibfield{author}{\bibinfo{person}{Andreas Zeller}, \bibinfo{person}{Rahul
  Gopinath}, \bibinfo{person}{Marcel B{\"o}hme}, \bibinfo{person}{Gordon
  Fraser}, {and} \bibinfo{person}{Christian Holler}.}
  \bibinfo{year}{2019}\natexlab{}.
\newblock \bibinfo{title}{The fuzzing book}.
\newblock
\newblock


\bibitem[Zhao et~al\mbox{.}(2019)]%
        {seqfuzzer}
\bibfield{author}{\bibinfo{person}{Hui Zhao}, \bibinfo{person}{Zhihui Li},
  \bibinfo{person}{Hansheng Wei}, \bibinfo{person}{Jianqi Shi}, {and}
  \bibinfo{person}{Yanhong Huang}.} \bibinfo{year}{2019}\natexlab{}.
\newblock \showarticletitle{SeqFuzzer: An Industrial Protocol Fuzzing Framework
  from a Deep Learning Perspective}. In \bibinfo{booktitle}{\emph{2019 12th
  IEEE Conference on Software Testing, Validation and Verification (ICST)}}.
  \bibinfo{pages}{59--67}.
\newblock
\urldef\tempurl%
\url{https://doi.org/10.1109/ICST.2019.00016}
\showDOI{\tempurl}


\bibitem[Zhao et~al\mbox{.}(2022)]%
        {zhao2022javatailor}
\bibfield{author}{\bibinfo{person}{Yingquan Zhao}, \bibinfo{person}{Zan Wang},
  \bibinfo{person}{Junjie Chen}, \bibinfo{person}{Mengdi Liu},
  \bibinfo{person}{Mingyuan Wu}, \bibinfo{person}{Yuqun Zhang}, {and}
  \bibinfo{person}{Lingming Zhang}.} \bibinfo{year}{2022}\natexlab{}.
\newblock \showarticletitle{History-Driven Test Program Synthesis for JVM
  Testing}. In \bibinfo{booktitle}{\emph{Proceedings of the 44th International
  Conference on Software Engineering}} (Pittsburgh, Pennsylvania)
  \emph{(\bibinfo{series}{ICSE '22})}. \bibinfo{pages}{1133–1144}.
\newblock


\bibitem[Zhou et~al\mbox{.}(2022)]%
        {zhou2022humanprompt}
\bibfield{author}{\bibinfo{person}{Yongchao Zhou}, \bibinfo{person}{Andrei~Ioan
  Muresanu}, \bibinfo{person}{Ziwen Han}, \bibinfo{person}{Keiran Paster},
  \bibinfo{person}{Silviu Pitis}, \bibinfo{person}{Harris Chan}, {and}
  \bibinfo{person}{Jimmy Ba}.} \bibinfo{year}{2022}\natexlab{}.
\newblock \showarticletitle{Large language models are human-level prompt
  engineers}.
\newblock \bibinfo{journal}{\emph{arXiv preprint arXiv:2211.01910}}
  (\bibinfo{year}{2022}).
\newblock


\bibitem[Ziegler et~al\mbox{.}(2019)]%
        {ziegler2019rlhf}
\bibfield{author}{\bibinfo{person}{Daniel~M. Ziegler}, \bibinfo{person}{Nisan
  Stiennon}, \bibinfo{person}{Jeffrey Wu}, \bibinfo{person}{Tom~B. Brown},
  \bibinfo{person}{Alec Radford}, \bibinfo{person}{Dario Amodei},
  \bibinfo{person}{Paul Christiano}, {and} \bibinfo{person}{Geoffrey Irving}.}
  \bibinfo{year}{2019}\natexlab{}.
\newblock \bibinfo{title}{Fine-Tuning Language Models from Human Preferences}.
\newblock
\newblock
\newblock
\shownote{arXiv:1909.08593}.


\end{thebibliography}

\end{document}